\documentclass[twocolumn,english,superscriptaddress,secnumarabic,amssymb,amsmath,nobibnotes,aps,prd,showkeys,showpacs]{revtex4}
\usepackage[latin9]{inputenc}
\usepackage{color}
\usepackage{verbatim}
\usepackage{textcomp}
\usepackage{amsmath}
\usepackage{amssymb}
\usepackage{graphicx}
\usepackage{esint}

\makeatletter
\@ifundefined{textcolor}{}
{%
 \definecolor{BLACK}{gray}{0}
 \definecolor{WHITE}{gray}{1}
 \definecolor{RED}{rgb}{1,0,0}
 \definecolor{GREEN}{rgb}{0,1,0}
 \definecolor{BLUE}{rgb}{0,0,1}
 \definecolor{CYAN}{cmyk}{1,0,0,0}
 \definecolor{MAGENTA}{cmyk}{0,1,0,0}
 \definecolor{YELLOW}{cmyk}{0,0,1,0}
}

\usepackage{epsfig}\usepackage{amstext}\usepackage{bm}%\documentclass[11pt,a4paper]{scrartcl}
\usepackage{booktabs}%%%%%%%%%%%%%%%%%%%%%%%%%%%%%%%%%%%%%%%%%%%%%%%%%%%%%%% some commands
\newcommand{\pd}[2]{\frac{\partial #1}{\partial #2}} 
\newcommand{\grad}[1]{\mathbf{\nabla} #1} % for gradient
\let\baraccent=\= % rename builtin command \= to \baraccent
\renewcommand{\=}[1]{\stackrel{#1}{=}} % for putting numbers above =

\usepackage{babel}

\usepackage{babel}

\usepackage{babel}

\usepackage{babel}

\usepackage{babel}

\makeatother

\usepackage{babel}
\begin{document}

\title{Comparing Hamiltonians of a spinning test particle for different
tetrad~fields}

\author{Daniela Kunst}

\email{daniela.kunst@zarm.uni-bremen.de}

\affiliation{ZARM, University of Bremen, Am Fallturm, 28359 Bremen, Germany}

\author{Tomáš Ledvinka}

\email{ledvinka@gmail.com}

\affiliation{Institute of Theoretical Physics, Faculty of Mathematics and Physics,
Charles University in Prague, 18000 Prague, Czech Republic}

\author{Georgios Lukes-Gerakopoulos}

\email{gglukes@gmail.com}

\affiliation{Institute of Theoretical Physics, Faculty of Mathematics and Physics,
Charles University in Prague, Czech Republic}

\affiliation{Theoretical Physics Institute, University of Jena, 07743 Jena, Germany}

\author{Jonathan Seyrich}

\email{seyrich@na.uni-tuebingen.de}

\affiliation{Mathematisches Institut, Universität Tübingen, Auf der Morgenstelle,
72076 Tübingen, Germany}
\begin{abstract}
This work is concerned with suitable choices of tetrad fields and
coordinate systems for the Hamiltonian formalism of a spinning particle
derived in {[}E. Barausse, E. Racine, and A. Buonanno,  \textit{Phys.
Rev. D} \textbf{80}, 104025 (2009){]}. After demonstrating that with
the originally proposed tetrad field the components of the total angular
momentum are not preserved in the Schwarzschild limit, we analyze
other hitherto proposed tetrad choices. Then, we introduce and thoroughly
test two new tetrad fields in the horizon penetrating Kerr--Schild
coordinates. Moreover, we show that for the Schwarzschild spacetime
background the linearized in spin Hamiltonian corresponds
to an integrable system, while for the Kerr spacetime we find chaos
which suggests a nonintegrable system. 
\end{abstract}


\maketitle

\section{Introduction}

\label{sec:Intro}

The motion of a spinning particle in the spacetime background of a
black hole, particularly the Kerr spacetime, is of great astrophysical
interest. Namely, it approximates the motion of a stellar compact
object (e.g., a black hole) around a supermassive black hole. Such
binary systems are expected to lie at the center of galaxies (see,
e.g., \cite{AmaroSeoane14} and references therein) and to be good
candidates for sources of gravitational radiation \cite{Riles13}.

Even though the equations describing the motion of a spinning particle
in a curved spacetime have been provided several decades ago by Mathisson
\cite{Mathisson37} and Papapetrou \cite{Papapetrou51}, many issues
of this motion are still open. The problem lies in the fact that the
Mathisson-Papapetrou (MP) equations are not a closed system of first
order differential equations. Hence, a spin supplementary condition
(SSC) is needed in order to close them. Several such SSCs have been
proposed (see, e.g., \cite{Semerak99,Kyrian07} for a review), each
of which introduces a different reference frame. Physically, the ambiguity
in the choice of a SSC is related to the fact that a spinning body
cannot be treated as a point particle but must have a finite size
in order to be prevented from rotating at superluminal speed \cite{Moeller49}.
In particular, each SSC corresponds to an observer who sees the reference
worldline fixed by the SSC as the center of mass of the extended body.
Previous studies have shown that the choice of the SSC depends on
the question one wants to investigate \cite{Costa12,Corinaldesi51,Kyrian07,Semerak99,LSK1,Moeller49,Pirani56,Semerak15}.

Within the MP equations the motion of spinning test particles in the
Schwarzschild or Kerr spacetime have been investigated in \cite{Hackmann14,Plyatsko12,Plyatsko13},
and several papers have been devoted to the investigation of the appearing
chaotic motion \cite{Hartl03a,Hartl03b,Suzuki97,Verhaaren10}. Beyond
the pole-dipole approximation, the quadrupole moment of the test particle
has already been taken into account \cite{Steinhoff12,Bini14}.

The dynamics of spinning test particles has not only been worked out in
Lagrangian formalisms (MP equations) \cite{Bailey75}, but in Hamiltonian
formalisms \cite{Steinhoff11,Steinhoff08,Barausse09,Tauber88,Deriglazov15a,Deriglazov15b}
as well. Hamiltonian dynamics has a long tradition in astronomy and a large 
number of problems there (e.g., perturbative problems or chaotic motion) are
typically studied from a Hamiltonian perspective \cite{Contop02}.
In general relativity the Hamiltonian formalisms have been applied, for example, in
the framework of the canonical Arnowitt-Deser-Misner (ADM) formalism
\cite{ADM} and in the effective-one-body approach (EOB) \cite{Buonanno99,Damour00,Damour08},
which studies the dynamics of spinning bodies of compact objects
using the Hamiltonian description of a one-body problem \cite{Blanchet06}.
Due to the significance of a Hamiltonian approach the Hamiltonian description of
spinning particles is important despite the fact that it mostly neglects terms
quadratic in spin.

In our previous work \cite{LSK1}, we have compared the Tulczyjew
(T) SSC \cite{Tulczyjew59} with the Newton-Wigner (NW) SSC \cite{NewtonWigner49}
as supplements to the MP equations. In a second step, we compared
the MP equations supplemented by the NW SSC to the corresponding Hamilton
equations derived in \cite{Barausse09} based on the same NW SSC.
In this work we focus on the latter, i.e., on a canonical Hamiltonian
formalism which should be equivalent to the MP equations up to the
linear order of the test particle spin.

In contrast to the T SSC, the NW SSC, which is used within the framework
of the Hamiltonian formalism, does not provide a unique choice of
reference frame. It rather defines an entire class of observers, each
characterized by a different tetrad field. Thus, the Hamiltonian formalism
proposed in \cite{Barausse09} depends on the choice of a reference
basis given by such a tetrad field. Each choice of a tetrad field
basically determines the form and the properties of the resulting
Hamiltonian function.
The fact that tetrads providing certain frames of reference are involved in a
definition of the spin variable can also be seen as a consequence of the fact
that in the Hamiltonian description the spin is a vector with prescribed canonical
relations to coordinates and momenta. Still, one might conclude that the tetrad
dependence of the Hamiltonian description of the spinning particle is against 
covariance principles of general relativity. Yet, when we numerically solve
equations of motion we have to use some coordinates anyway. The involvement of 
tetrads simply means we use different coordinates for external and inner degrees
of freedom of the spinning particle. As, e.g., Boyer-Lindquist coordinates are
comfortable for solving equations of motion in Kerr geometry there may well be
some other tetrad fields more suitable for the definition of the Hamiltonian spin.

We discuss the advantages and the drawbacks
of Hamiltonian functions arising from tetrad fields already proposed
in \cite{Barausse09,Barausse10}. Then, we introduce two new tetrad
fields in Kerr--Schild coordinates which yield Hamiltonian functions
with desirable properties using both analytical and
numerical analysis. Namely in order to have a good choice of a tetrad
field, the corresponding Hamiltonian should reflect the symmetries
of the background spacetime, i.e., preserve the integrals of motion,
and avoid any coordinate effects evoked by coordinate dependent tetrad
basis vectors.
For the above discussion, we focus on the Schwarzschild limit and
show that the well behaving Hamiltonian functions based on our tetrads
have as many integrals of motion as degrees of freedom.
Thus, it is shown that in the Schwarzschild limit these Hamiltonians describe
an integrable system.
We view this as an important test, as in general, for different tetrad fields
the description \cite{Barausse09} provides Hamiltonians non-equivalent beyond the
given approximation. In any Hamiltonian system the integrals of
motion play a crucial role, when the integrability issue is studied.
If we have several possible descriptions of the same system in the given
approximation, those respecting all background symmetries are the obvious choice.
We use the case of spinning particle in Schwarzschild spacetime as such an
exact problem with many integrals of motion to demonstrate shortcomings of
certain coordinate-tetrad choices. Even though the considered approximations
assume small spins, to clearly demonstrate (non-)integrability we also use large
spin values in numerical tests.

As for the Kerr spacetime, it was shown in~\cite{Rudiger} that if
the MP equations supplemented by the T SSC are linearized in the spin,
an integral of motion associated with a Killing-Yano tensor appears.
This led to the impression that, up to linear order in the spin, the
motion of a spinning particle is integrable in general \cite{Hinderer13}.
However, according to our numerical calculations, this seems not to
be the case for the Hamiltonian function depending
on the tetrad field choice introduced in \cite{Barausse10}.

This paper is organized as follows. In Sec. \ref{sec:HamForm} we
give a short overview of the Hamiltonian formalism introduced in \cite{Barausse09}.
After that, in Sec. \ref{sec:HamBL}, we present two different choices
of coordinate systems, the Boyer-Lindquist and the Cartesian isotropic coordinates,
for a tetrad corresponding to a ZAMO observer which is already given
in \cite{Barausse09,Barausse10}. We analyze the properties of both
with the help of analytical calculations and numerical integrations.
Then we present our new tetrads in Kerr-Schild coordinates in Sec.
~\ref{sec:RHamKS}. Finally, in Sec. ~\ref{sec:Conclusions}, we
summarize our results. We add a description of our numerical tools
in the Appendix.

We use geometric units, i.e., $(G=c=1)$, and the signature of the
metric is (-,+,+,+). Greek letters denote the indices corresponding
to spacetime (running from 0 to 3), while Latin letters denote indices
corresponding only to space (running from 1 to 3).

\section{The Hamiltonian formalism}

\label{sec:HamForm}

The Hamiltonian formalism in \cite{Barausse09} has been achieved
by linearizing the MP equations of motion for the NW SSC. The MP equations
describe the motion of a particle with mass $m^{2}=-p_{\mu}p^{\mu}$,
satisfying the mass shell constraint, and spin $S^{\mu\nu}$ in a
given spacetime background $g_{\mu\nu}$. Their reformulation in \cite{Dixon70}
reads 
\begin{eqnarray}
\frac{D~p^{\mu}}{d\sigma}=-\frac{1}{2}~{R^{\mu}}_{\nu\kappa\lambda}v^{\nu}S^{\kappa\lambda}~~,\label{eq:MPmomenta}\\
\nonumber \\
\frac{D~S^{\mu\nu}}{d\sigma}=p^{\mu}~v^{\nu}-v^{\mu}~p^{\nu}~~,\label{eq:MPspin}
\end{eqnarray}
where $p^{\mu}$ is the four-momentum, $v^{\mu}=dx^{\mu}/d\sigma$
is the tangent vector to the worldline along which the particle moves,
$\sigma$ is an evolution parameter along this worldline, and ${R^{\mu}}_{\nu\kappa\lambda}$
is the Riemann tensor. The NW SSC reads 
\begin{equation}
S^{\mu\nu}~\omega_{\mu}=0~~,\label{eq:NW}
\end{equation}
where $\omega_{\mu}$ is a sum of timelike vectors. This sum in \cite{Barausse09}
has the form 
\begin{equation}
\omega_{\nu}=p_{\nu}-m~\tilde{e}_{\nu}\,^{T}~~,\label{eq:NWTimeLV}
\end{equation}
where $\tilde{e}_{\nu}\,^{T}$ is a timelike future
oriented vector (throughout the article we use T instead of 0), which
together with three spacelike vectors $\tilde{e}^{\mu}\,_{I}$, denoted
by capital latin indices, is part of a tetrad field $\tilde{e}^{\mu}\,_{\Delta}$.

This tetrad field has to satisfy two conditions: the first condition
ensures the orthonormality of the tetrad given by 
\begin{equation}
\tilde{e}^{\mu}\,_{\Gamma}\tilde{e}^{\nu}\,_{\Delta}~g_{\mu\nu}=\eta_{\Gamma\Delta}~~,\label{eq:tetradflat}
\end{equation}
where $\eta_{\Gamma\Delta}$ is the metric of the flat spacetime and
$g_{\mu\nu}$ its analogon for the curved spacetime background. The
capital indices are raised and lowered by the flat metric $\eta_{\Gamma\Delta}$,
the small ones by $g_{\mu\nu}$. The second condition is implied by
\eqref{eq:tetradflat} and reads 
\begin{equation}
\tilde{e}^{\mu}\,_{\Delta}\tilde{e}_{\nu}\,^{\Delta}=\delta_{\nu}^{\mu}~~,\label{eq:tetradcom}
\end{equation}
where $\delta_{\nu}^{\mu}$ is the Kronecker delta.

When a tensor is projected on the tetrad field, then it is denoted
with capital indices. For example, $\omega_{\Delta}=\tilde{e}^{\nu}\,_{\Delta}\omega_{\nu}$
is the projection of the time-like vector (\ref{eq:NWTimeLV}) on
the tetrad field, i.e., 
\begin{eqnarray}
\omega_{T} & = & p_{\nu}~\tilde{e}^{\nu}\,_{T}-m~~,\nonumber \\
\omega_{J} & = & p_{\nu}~\tilde{e}^{\nu}\,_{J}~~.\label{eq:omega}
\end{eqnarray}

Then, the spin tensor $S^{\mu\nu}$ projection reads
\begin{equation}
S^{IJ}=S^{\mu\nu}~\tilde{e}_{\mu}\,^{I}~\tilde{e}_{\nu}\,^{J}~~.\label{eq:SpinProj}
\end{equation}
In~\cite{Barausse09}, the authors do not work with this tensor but
rather employ the spin three vector 
\begin{equation}
S_{I}=\frac{1}{2}\epsilon_{IJL}~S^{JL}~~,\label{eq:3ProjSpin}
\end{equation}
where $\epsilon_{IJL}$ is the Levi-Civita symbol.

Now, the Hamiltonian function $H$ for a spinning particle 
\begin{equation}
H=H_{NS}+H_{S}~~,\label{eq:HamSP}
\end{equation}
splits in two parts. The first, 
\begin{equation}
H_{NS}=\beta^{i}P_{i}+\alpha~\sqrt{m^{2}+\gamma^{ij}P_{i}P_{j}}~~,\label{eq:HamNSP}
\end{equation}
is the Hamiltonian for a nonspinning particle, where 
\begin{eqnarray}
\alpha & = & \frac{1}{\sqrt{-g^{00}}}~~,\label{eq:alpha}\\
\beta^{i} & = & \frac{g^{0i}}{g^{00}}~~,\label{eq:beta}\\
\gamma^{ij} & = & g^{ij}-\frac{g^{0i}g^{0j}}{g^{00}}~~,\label{eq:gamma}
\end{eqnarray}
and $P_{i}$ are the canonical momenta conjugate to $x^{i}$ of the
Hamiltonian~\eqref{eq:HamSP}. They can be calculated from the momenta
$p_{i}$ with the help of the relation 
\begin{eqnarray}
P_{i} & = & p_{i}+E_{i\Gamma\Delta}S^{\Gamma\Delta}~~,\nonumber \\
 & = & p_{i}+\left(2E_{iTJ}\frac{\omega_{C}}{\omega_{T}}+E_{iJC}\right)\epsilon^{JCL}~S_{L}~~,\label{eq:momentaHL}
\end{eqnarray}
where the spin connection 
\begin{equation}
E_{\nu\Gamma\Delta}=-\frac{1}{2}\left(g_{\kappa\lambda}~\tilde{e}^{\kappa}\,_{\Gamma}~\frac{\partial\tilde{e}^{\lambda}\,_{\Delta}}{\partial x^{\nu}}+\tilde{e}^{\kappa}\,_{\Gamma}~\Gamma_{\kappa\nu\lambda}~\tilde{e}^{\lambda}\,_{\Delta}\right)~~,\label{eq:EmuAB}
\end{equation}
is a tensor which is antisymmetric in the last two indices, i.e.,
$E_{\nu\Gamma\Delta}=-E_{\nu\Delta\Gamma}$, and $\Gamma_{\kappa\nu\lambda}$
are the Christoffel symbols. The second part of the Hamiltonian, 
\begin{equation}
H_{S}=-\left(\beta^{i}F_{i}^{C}+F_{0}^{C}+\frac{\alpha~\gamma^{ij}P_{i}~F_{j}^{C}}{\sqrt{m^{2}+\gamma^{ij}P_{i}P_{j}}}\right)S_{C}~~,\label{eq:HamSPc}
\end{equation}
provides the contribution of the particle's spin to the motion, with
\begin{equation}
F_{\mu}^{C}=\left(2E_{\mu TI}\frac{\bar{\omega}_{J}}{\bar{\omega}_{T}}+E_{\mu IJ}\right)\epsilon^{IJC}~~,\label{eq:FmuC}
\end{equation}
and 
\begin{eqnarray}
\bar{\omega}_{\Delta} & = & \bar{\omega}_{\nu}~\tilde{e}^{\nu}\,_{\Delta}~~,\nonumber \\
\bar{\omega}_{\nu} & = & \bar{P}_{\nu}-m~\tilde{e}_{\nu}\,^{T}~~,\nonumber \\
\bar{P}_{i} & = & P_{i}~~,\nonumber \\
\bar{P}_{0} & = & -\beta^{i}~P_{i}-\alpha\sqrt{m^{2}+\gamma^{ij}P_{i}P_{j}}~~,\nonumber \\
\bar{\omega}_{T} & = & \bar{P}_{\nu}~\tilde{e}^{\nu}\,_{T}-m~~,\nonumber \\
\bar{\omega}_{J} & = & \bar{P}_{\nu}~\tilde{e}^{\nu}\,_{~J}~~.\label{eq:omegabar}
\end{eqnarray}

The equations of motion for the canonical variables as a function
of coordinate time $t$ read 
\begin{align}
\frac{dx^{i}}{dt} & =\pd H{P_{i}}~~,\label{eqn-Ham-eq-of-motion-x}\\
\frac{dP_{i}}{dt} & =-\pd H{x^{i}}~~,\label{eqn-Ham-eq-of-motion-P}\\
\frac{dS_{I}}{dt} & =\epsilon_{IJC}\pd H{S_{J}}S^{C}~~~~.\label{eqn-Ham-eq-of-motion-S}
\end{align}
The phase space of a canonical Hamiltonian system
is equipped with a binary operation, i.e., the Poisson bracket. If
the dynamical system is subject to (secondary) constraints $\xi_{i}$,
the Poisson bracket has to be replaced by the Dirac bracket \cite{Dirac}
\begin{equation}
\left\{ Q,R\right\} _{DB}:=\left\{ Q,R\right\} -\left\{ Q,\xi_{i}\right\} \left[C^{-1}\right]_{ij}\left\{ \xi_{j},R\right\} \,\,,\label{eq:DiracBracketDef}
\end{equation}
where $Q$ and $R$ are functions on phase space and $C^{-1}$ is the
inverse of the matrix consisting of the Poisson brackets of the set
of constraints $C=\left\{ \xi_{i},\xi_{j}\right\} $. In the case
of a spinning particle the constraints are given by the supplementary
condition, here the NW SSC \eqref{eq:NW}, and, in order to retain
the symplectic structure, by the choice of the timelike body-fixed
tetrad vector to be aligned with the four momentum 
\[
\chi_{\mu}:=e_{\mu T}-\frac{p_{\mu}}{m}=0\,\,,
\]
where $e^{\mu}\,_{A}$ is related to the local frame $\tilde{e}^{\mu}\,_{A}$
by a Lorentz transformation (for more details see
\cite{Barausse09}). In order to derive the canonical structure of
the phase space variables, the new defined momenta $P_{\mu}$ in eq.
\eqref{eq:momentaHL} are treated as functions of the kinematical
momenta $p_{\mu}$, the position, and of the spin which result in
the following bracket relations 
\begin{align}
\left\{ x^{i},P_{j}\right\} _{DB} & =\delta_{j}^{i}+\mathcal{O}\left(S^{2}\right)~~,\nonumber \\
\left\{ S^{I},S^{J}\right\} _{DB} & =\epsilon^{IJK}S^{K}+\mathcal{O}\left(S^{2}\right)~~.\label{eq:PhaseSpaceStructure}
\end{align}
All the other bracket relations between the variables
vanish at linear order in spin \cite{Barausse09}. At this approximation, even if
the mass $m^{2}=-p_{\nu}p^{\nu}$ is not a constant of motion for the exact MP
equations with NW SSC, actually it scales quadratically in the particle's spin
(see, e.g., \cite{LSK1}), the mass is preserved at first order in the spin and 
treated as a constant in the linearized Hamiltonian formalism \cite{Barausse09}.

When we restrict the scheme to the linearized Hamiltonian
formalism, and consider the $P_{i}$ no longer as functions but as
independent phase space variables, then the terms of $\mathcal{O}\left(S^{2}\right)$
are dropped in all the above Dirac brackets, i.e., in \eqref{eq:PhaseSpaceStructure}
and all the other bracket relations between the variables $\{x^{i},P_{i},S_{I}\}$.
Profoundly, in the linearized Hamiltonian formalism a quantity $I$
is a constant of motion, if it holds
for its Dirac bracket with the Hamiltonian function $H$ 
\begin{equation}
\left\{ I,H\right\} _{DB}=0~~.
\end{equation}
This means that if the system is evolved by the eqs. \eqref{eqn-Ham-eq-of-motion-x}-\eqref{eqn-Ham-eq-of-motion-S},
then the quantity $I$ is preserved during the evolution.

The formulation provided up to this point is general, namely it does
not depend on the specific coordinate system or on the specific tetrad
field. These two factors, however, are essential for the Hamiltonian
function~\eqref{eq:HamSP}. In particular, the nonspinning part of
the Hamiltonian function~\eqref{eq:HamNSP} depends on the coordinate
system which the metric is written in, while the spinning part~\eqref{eq:HamSPc}
depends on the tetrad we choose. In Sec.~\ref{sec:HamBL} and Sec.~\ref{sec:RHamKS},
we present three different combinations tetrad $\leftrightarrow$
coordinates for the Kerr spacetime background and discuss the advantages
and shortcomings of the respective setups.

\section{The Hamiltonian Function in Boyer-Lindquist coordinates compared
with Cartesian Isotropic coordinates}

\label{sec:HamBL}

\subsection{A tetrad in Boyer-Lindquist coordinates}

\label{sec:OHamBL}

A Hamiltonian function for the Kerr spacetime background in Boyer-Lindquist
(BL) has been provided in \cite{Barausse09}. The line element of
the Kerr spacetime in BL coordinates reads 
\begin{eqnarray}
ds^{2} & = & g_{tt}~dt^{2}+2~g_{t\phi}~dt~d\phi+g_{\phi\phi}~d\phi^{2}\nonumber \\
 & + & g_{rr}~dr^{2}+g_{\theta\theta}~d\theta^{2}~~,\label{eq:LinEl}
\end{eqnarray}
with 
\begin{eqnarray}
g_{tt} & = & -1+\frac{2Mr}{\Sigma}~~,\nonumber \\
g_{t\phi} & = & -\frac{2aMr\sin^{2}{\theta}}{\Sigma}~~,\nonumber \\
g_{\phi\phi} & = & \frac{\Lambda\sin^{2}{\theta}}{\Sigma}~~,\label{eq:KerrMetric}\\
g_{rr} & = & \frac{\Sigma}{\Delta}~~,\nonumber \\
g_{\theta\theta} & = & \Sigma~~,\nonumber 
\end{eqnarray}
and 
\begin{eqnarray}
\Sigma & = & r^{2}+a^{2}\cos^{2}{\theta}~~,\nonumber \\
\Delta & = & \varpi^{2}-2Mr~~,\nonumber \\
\varpi^{2} & = & r^{2}+a^{2}~~,\nonumber \\
\Lambda & = & \varpi^{4}-a^{2}\Delta\sin^{2}\theta~~.\label{eq:Kerrfunc}
\end{eqnarray}
$M$ denotes the mass and $a$ the spin parameter of the central Kerr
black hole.

The tetrad field given in \cite{Barausse09} reads 
\begin{eqnarray}
\tilde{e}_{\mu}\,^{T} & = & \delta_{\mu}^{t}\sqrt{\frac{\Delta\Sigma}{\Lambda}}~~,\nonumber \\
\tilde{e}_{\mu}\,^{1} & = & \delta_{\mu}^{r}\sqrt{\frac{\Sigma}{\Delta}}~~,\nonumber \\
\tilde{e}_{\mu}\,^{2} & = & \delta_{\mu}^{\theta}\sqrt{\Sigma}~~,\nonumber \\
\tilde{e}_{\mu}\,^{3} & = & -\delta_{\mu}^{t}\frac{2aMr\sin\theta}{\sqrt{\Lambda\Sigma}}+\delta_{\mu}^{\phi}\sin{\theta}\sqrt{\frac{\Lambda}{\Sigma}}~~,\label{eq:TetradBL}
\end{eqnarray}
where for the small indices the numbers have been replaced with the
corresponding coordinates, i.e., $t,~r,~\theta,~\phi$ stand for $0,~1,~2,~3$,
respectively. The proposed tetrad corresponds to an observer in the
zero angular momentum frame (ZAMO) which intuitively yields a reasonable
choice. Moreover, the coordinate system is based on the spherical
coordinates in flat spacetime which respects the symmetries of the
spacetime. In the Schwarzschild limit the above tetrad field reduces
to $(a\rightarrow0)$ 
\begin{eqnarray}
\tilde{e}_{\mu}\,^{T} & = & \delta_{\mu}^{t}\sqrt{f(r)}~~,\nonumber \\
\tilde{e}_{\mu}\,^{1} & = & \delta_{\mu}^{r}\sqrt{f(r)^{-1}}~~,\nonumber \\
\tilde{e}_{\mu}\,^{2} & = & r~\delta_{\mu}^{\theta}~~,\nonumber \\
\tilde{e}_{\mu}\,^{3} & = & r~\sin{\theta}~\delta_{\mu}^{\phi}~~,\label{eq:TetradSchwSph}
\end{eqnarray}
where $f(r)=1-\frac{2M}{r}$. In the flat spacetime limit ($M\rightarrow0,\, a\rightarrow0$)
we get 
\begin{eqnarray}
\tilde{e}_{\mu}\,^{T} & = & \delta_{\mu}^{t}~~,\nonumber \\
\tilde{e}_{\mu}\,^{1} & = & \delta_{\mu}^{r}~~,\nonumber \\
\tilde{e}_{\mu}\,^{2} & = & r~\delta_{\mu}^{\theta}~~,\nonumber \\
\tilde{e}_{\mu}\,^{3} & = & r~\sin{\theta}~\delta_{\mu}^{\phi}~~.
\end{eqnarray}
This yields the flat spacetime in spherical coordinates.

Let's have a closer look at the dynamics in Schwarzschild spacetime.
The corresponding metric in Schwarzschild spacetime results in 
\[
ds^{2}=-f\left(r\right)dt^{2}+f\left(r\right)^{-1}dr^{2}+r^{2}\left(d\theta^{2}+\sin^{2}\left(\theta\right)d\phi^{2}\right)~~,
\]
with $f\left(r\right)=1-2M/r$ and the corresponding tetrad field
is \eqref{eq:TetradSchwSph}. The Hamiltonian 
\[
H=H_{NS}+H_{S}~~,
\]
is expressed in terms of the new phase space variables $\left(r,\theta,\phi,P_{r},P_{\theta},P_{\phi},S_{I}^{BL}\right)$
where $S_{I}^{BL}$ stands for the spin projected onto the spatial
background tetrad in spherical coordinates (reduced from the Boyer-Lindquist
coordinates). All told, we have 
\begin{align}
H= & \frac{1}{\sqrt{f\left(r\right)}}\sqrt{Q}+\frac{M}{r^{3}\left(1+\sqrt{Q}\right)}\left(P_{\theta}S_{3}^{BL}-\frac{P_{\phi}}{\sin\left(\theta\right)}S_{2}^{BL}\right)\nonumber \\
 & -\frac{f\left(r\right)}{r^{2}\sqrt{Q}}\bigg(\frac{\cos\left(\theta\right)}{\sin^{2}\theta\sqrt{f\left(r\right)}}P_{\phi}S_{1}^{BL}-\frac{P_{\phi}}{\sin\left(\theta\right)}S_{2}^{BL}\nonumber \\
 & +P_{\theta}S_{3}^{BL}\bigg)~~,\label{eq:Hspherical}
\end{align}
where $Q=m^{2}+f\left(r\right)P_{r}^{2}+\frac{1}{r^{2}}P_{\theta}^{2}+\frac{1}{r^{2}\sin^{2}\left(\theta\right)}P_{\phi}^{2}$.

In \cite{Barausse09} a criterion for the behavior of the Hamiltonian in the flat
spacetime limit was introduced in order to check whether the choice of
coordinates is a ``good'' one. Ideally, the contributions from the spin to the
Hamiltonian $H_{S}$ vanish, since we no longer have curvature which the spin
could couple to and the trajectory of the spinning particle should
simply be the one of a straight line. Thus, the motion of the particle
should be completely independent of the spin. However, in the case
of spherical coordinates the Hamiltonian is given by \eqref{eq:Hspherical}
and the contribution from the spin part $H_{S}$ does not vanish representing
an evolution of the spin in the absence of spin-orbit coupling, as
was noted in \cite{Barausse09}, which might imply a coordinate effect
for this choice of tetrad. Following the latter line of thought, we might say 
that the basis vectors are coordinate dependent, since they are oriented along
the direction of the coordinate basis vectors in spherical coordinates.
Therefore, they introduce an additional evolution to the dynamical system which
affects the equations of motion for the spinning particle, i.e., the equations
of motion do not only contain the physical dynamics of the spinning object, but
also the coordinate dynamics. On the other hand, the coordinate effect might not 
be the only interpretation, for instance for a long time the helical motion of a
spinning particle with Pirani SSC in the flat spacetime was considered unnatural,
until it was explained in terms of a hidden momentum in \cite{Costa12}. 
Anyhow, such effects make it harder to gain insights into the physical behavior
of the particle's motion, since it is not so easy to distinguish between
coordinate effects and physical effects in the results. Therefore, we prefer 
to focus on a more solid criterion for the Hamiltonian to check whether the
choice of coordinates is a ``good'' one, and this criterion comes from the
symmetries of the system.

\begin{figure}[htp]
\centerline{ \includegraphics[width=0.25\textwidth]{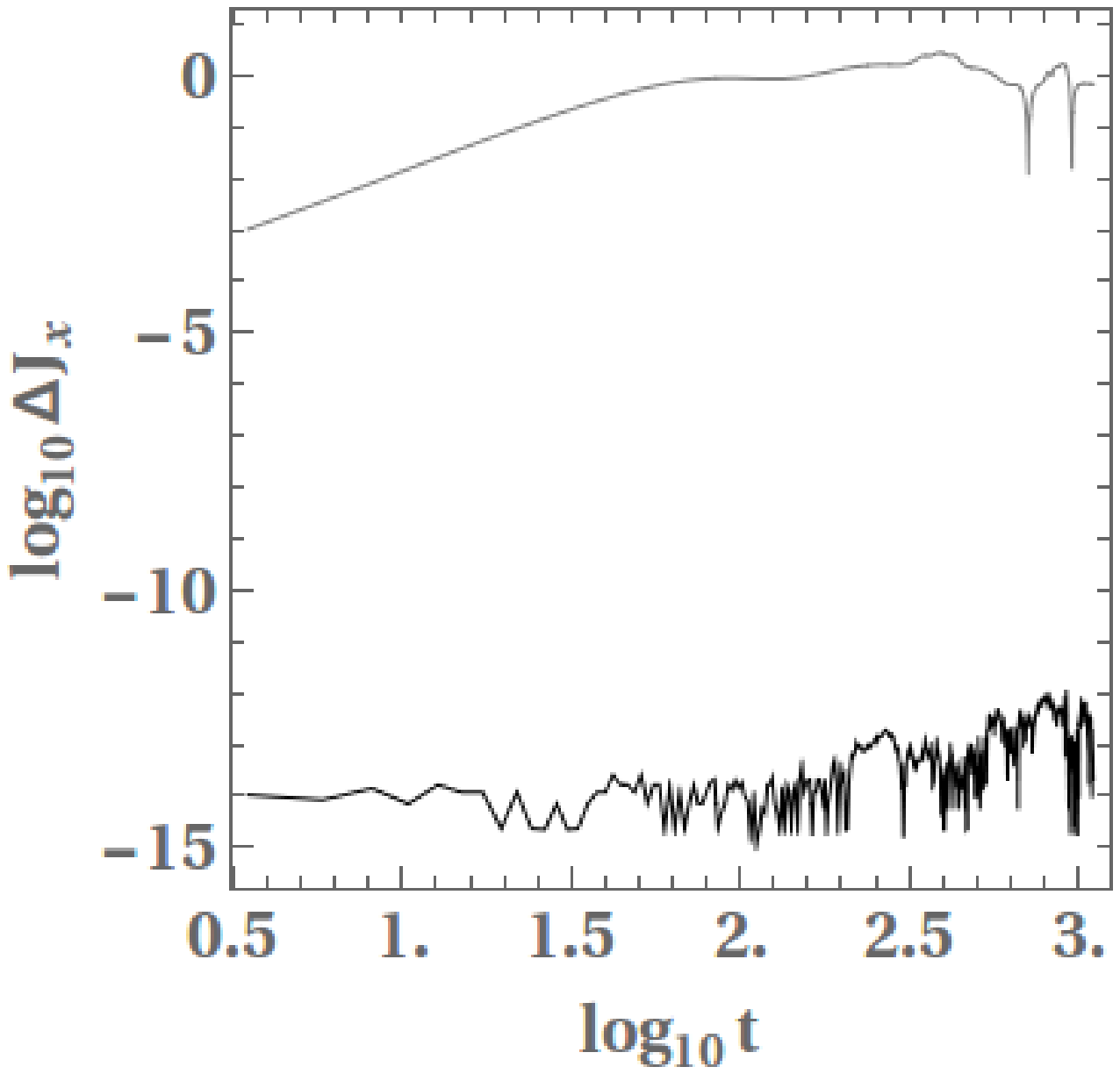} \includegraphics[width=0.25\textwidth]{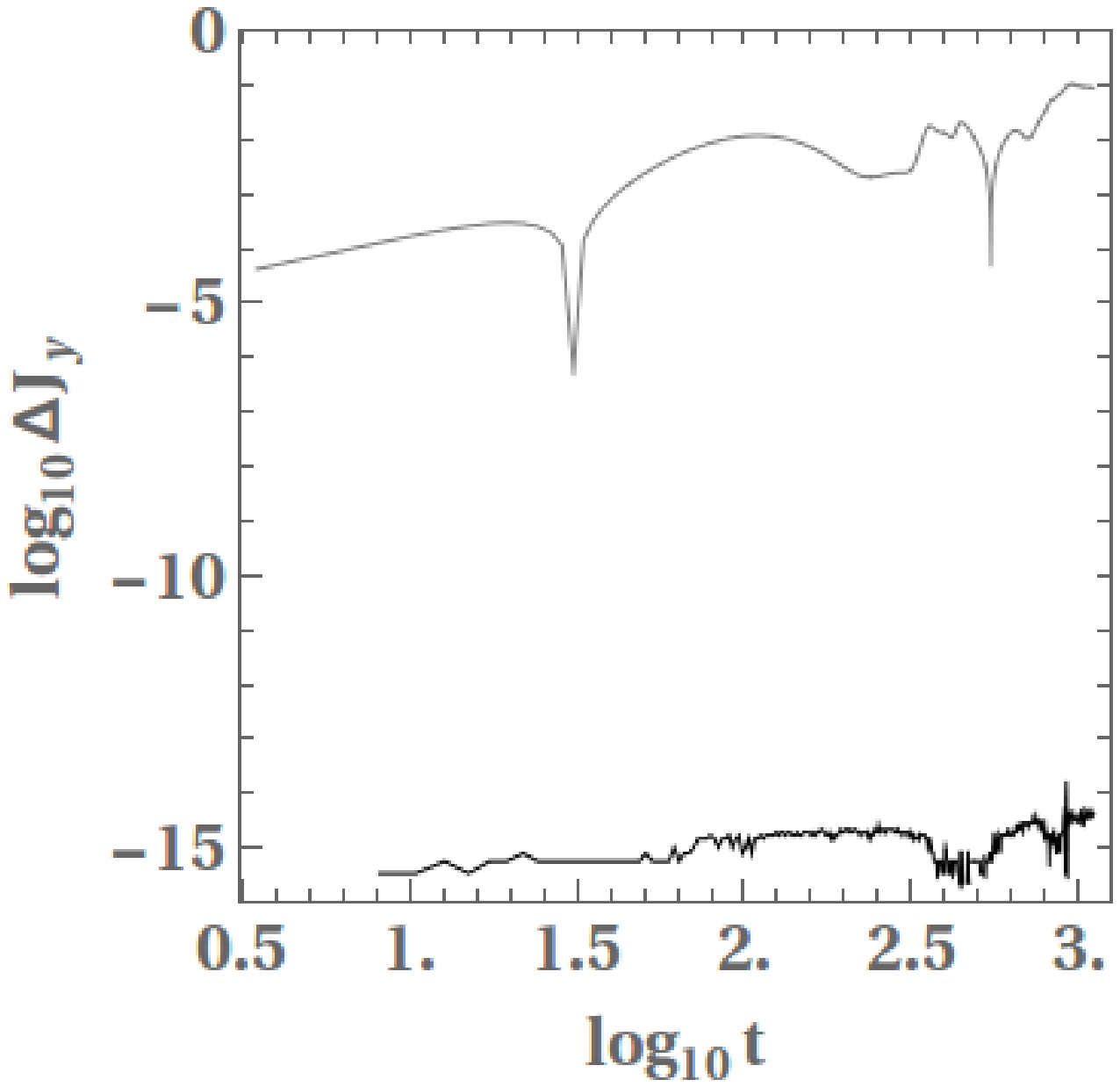}}
\caption{The left panel shows the relative error of $\Delta J_{x}$, and the
right of $\Delta J_{y}$ as a function of time in logarithmic scale
for the Schwarzschild background. The gray lines show the relative
error of these quantities when the system is evolved using the Hamiltonian
function corresponding to the tetrad \eqref{eq:TetradSchwSph}, while
the black lines show the relative error of these quantities when the
system is evolved using the respective MP equations. Both evolutions
share the same initial conditions conditions, where $a=0$, $M=m=1$,
and $S=1$.}
\label{Fig:DH1DS2ScHa05-1} 
\end{figure}

\begin{figure}[htp]
\centerline{ \includegraphics[width=0.25\textwidth]{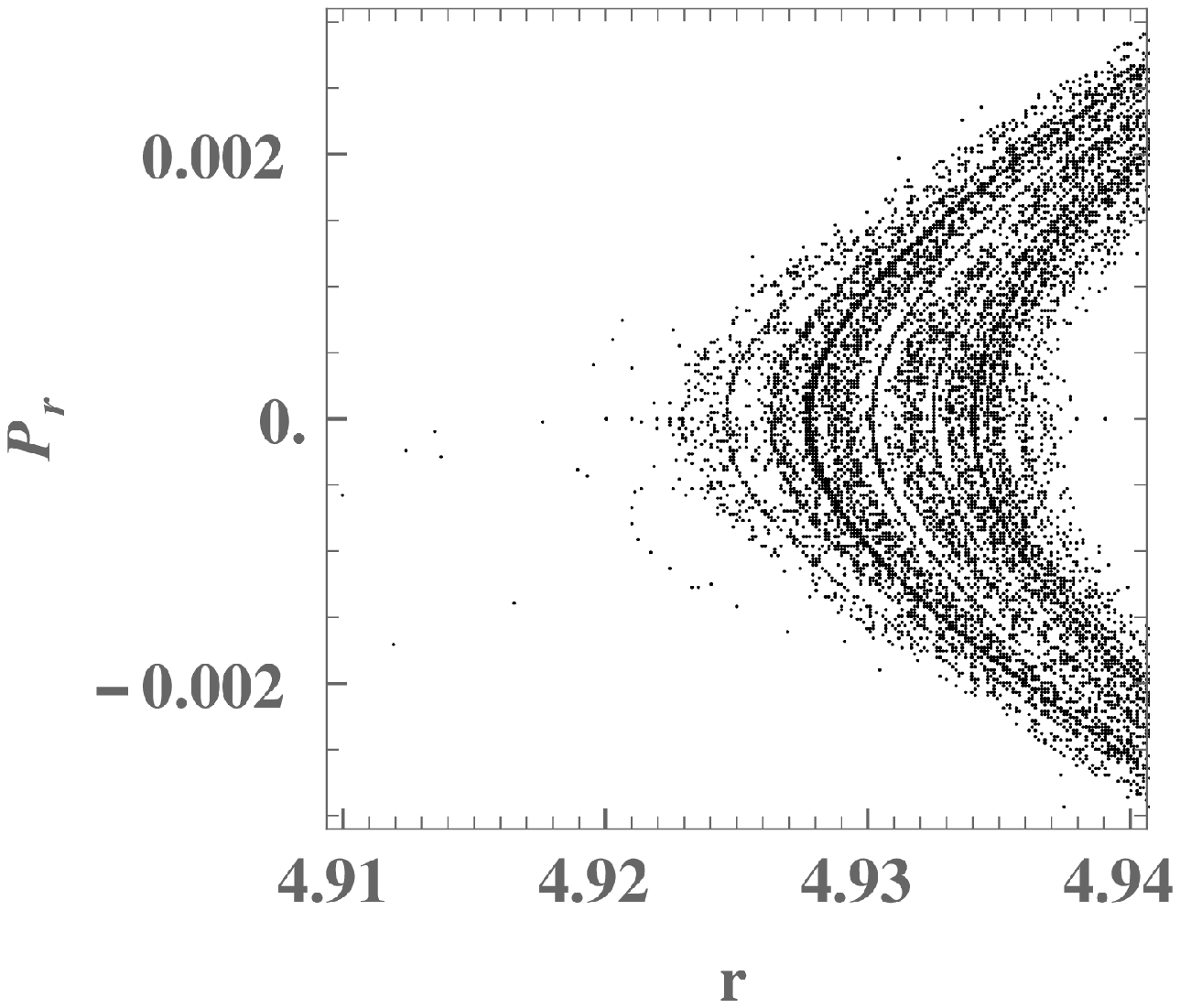} \includegraphics[width=0.25\textwidth]{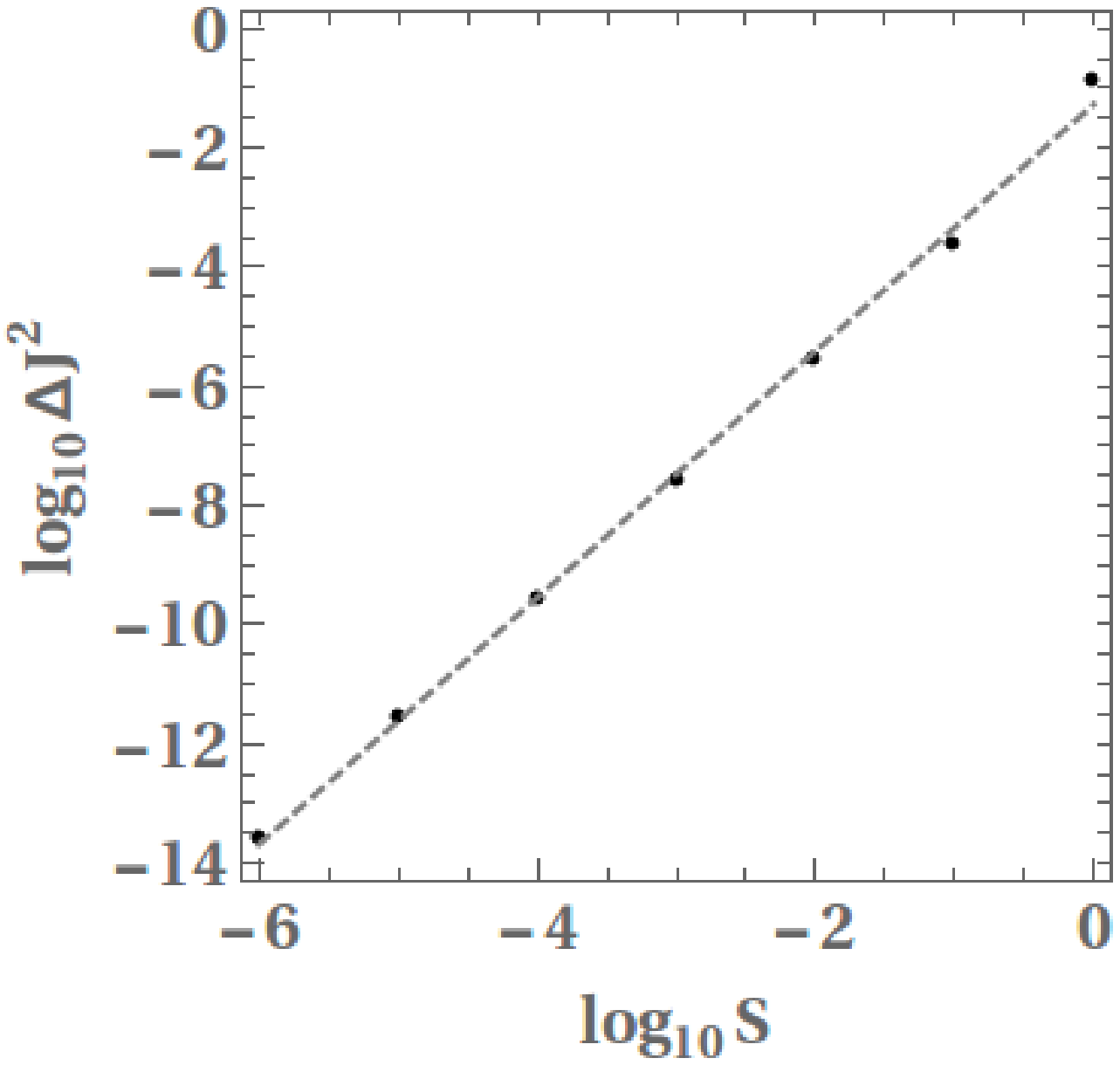}}
\caption{In  the left panel is a detail from the surface of section $\theta=\pi/2$,
$P_{\theta}>0$. The parameters of the orbits are $H=0.95$, $J_{z}=P_\phi=3$,
$S=M=m=1$, $a=0$, the common initial conditions are $\phi=0$,
$P_{r}=0$, $S_{1}=S_{2}=0$, $S=S_{3}=10^{-2}$, while the initial $P_\theta$ is
defined numerically by the Hamiltonian function $H$. The right panel shows how 
the maximal value of the relativity error $\Delta J^2=|1-J^2(t)/J^2(0)|$ for 
evolution intervals $t=10^3$ scales with the spin measure $S$. 
}
\label{Fig:DJ2SchCh} 
\end{figure}

Generally, according to Noether's theorem each spacetime symmetry
is related to a conserved quantity. In the case of spinning particles
moving in a particular spacetime geometry equipped with a symmetry
described by a Killing vector $\xi_{\mu}$, the associated quantity
conserved by MP equations reads 
\begin{equation}
C=p^{\mu}\xi_{\mu}-\frac{1}{2}S^{\mu\nu}\xi_{\mu;\nu}~~.\label{eq:conservedC}
\end{equation}
In Schwarzschild spacetime we have three spatial Killing vectors yielding
the three components of the total angular momentum \cite{Suzuki97}
\begin{align*}
J_{x}= & -p_{\theta}\sin\left(\phi\right)-p_{\phi}\cot\left(\theta\right)\cos\left(\phi\right)\\
 & +r^{2}S^{\theta\phi}\sin\left(\theta\right)^{2}\cos\left(\phi\right)+rS^{\phi r}\sin\left(\theta\right)\cos\left(\theta\right)\cos\left(\phi\right)\\
 & -rS^{r\theta}\sin\left(\phi\right)~~,\\
J_{y}= & p_{\theta}\cos\left(\phi\right)-p_{\phi}\cot\left(\theta\right)\sin\left(\phi\right)+r^{2}S^{\theta\phi}\sin\left(\theta\right)^{2}\sin\left(\phi\right)\\
 & +rS^{\phi r}\sin\left(\theta\right)\cos\left(\theta\right)\sin\left(\phi\right)+rS^{r\theta}\cos\left(\phi\right)~~,\\
J_{z}= & p_{\phi}-r\sin\left(\theta\right)^{2}\left(S^{\phi r}-rS^{\theta\phi}\cot\left(\theta\right)\right)~~,
\end{align*}
where $p_{i}$ are the kinematical momenta and $S^{ij}$ the spin
components written in coordinate basis. In order to check whether
the components of the total angular momentum are constants of motion
within the Hamiltonian formulation we have to transform the expression
to canonical variables $P_{i}$ and $S_{I}^{BL}$ with the relations
given in \eqref{eq:SpinProj} and \eqref{eq:momentaHL}. Therewith
we obtain 
\begin{align*}
J_{x} & =\cos(\phi)(S_{1}^{BL}\csc(\theta)-P_{\phi}\cot(\theta))-P_{\theta}\sin(\phi)~~,\\
J_{y} & =P_{\theta}\cos(\phi)+\sin(\phi)(S_{1}^{BL}\csc(\theta)-P_{\phi}\cot(\theta))~~,\\
J_{z} & =P_{\phi}~~,
\end{align*}
for the components of the total angular momentum, with which we may
now compute the evolution equations for $J_{i}$ via the Dirac brackets
with the Hamiltonian. Indeed, they result in 
\begin{align*}
\left\{ J_{x},H\right\} _{DB} & =\mathcal{O}\left(S^{2}\right)~~,\\
\left\{ J_{y},H\right\} _{DB} & =\mathcal{O}\left(S^{2}\right)~~,\\
\left\{ J_{z},H\right\} _{DB} & =0~~,
\end{align*}
and 
\begin{align*}
\left\{ J_{x}^{2}+J_{y}^{2}+J_{z}^{2},H\right\} _{DB} & =\mathcal{O}\left(S^{2}\right)~~.
\end{align*}
Although we consistently keep the linearization in the Hamiltonian
and the corresponding bracket structure, we find that the Dirac brackets
for $J_{x}$, $J_{y}$ and $J_{z}$ contain contributions from higher
orders in the particle's spin. Indeed, $J_{x}$ and $J_{y}$ start
oscillating when the Hamiltonian system corresponding to the tetrad
field \eqref{eq:TetradSchwSph} is numerically evolved through the
the equations of motion \eqref{eqn-Ham-eq-of-motion-x}-\eqref{eqn-Ham-eq-of-motion-S}.
It is visible from the relative error 
\begin{align}
\Delta J_{i}=|1-\frac{J_{i}(t)}{J_{i}(0)}|~~~i=x,y~~,\label{eq:ErrorJ}
\end{align}
at time $t$ of the $J_{x}$ and $J_{y}$ (gray line) in Fig.~\ref{Fig:DH1DS2ScHa05-1},
that the Hamiltonian function resulting from the tetrad \eqref{eq:TetradSchwSph}
apparently violates the symmetry properties of the Schwarzschild spacetime.
Consequently, the total angular momentum $J^{2}$ is not preserved,
because the $x$ and $y$ components of the total angular momentum
exhibit inappropriate behavior. On the other hand, the respective
evolution using the MP equation supplemented with NW SSC, instead,
shows the expected preservation of the angular momentum components
(black curves in Fig.~\ref{Fig:DH1DS2ScHa05-1}). This shows that
even in the above linear in spin Hamiltonian approximation a quantity
is a constant of motion only when its Dirac brackets with the Hamiltonian
are exactly zero, while when the brackets have contributions from
the higher in spin orders, the quantities show no constancy. The violation of the expected symmetries results in a system that exhibits
chaotic motion (scattered dots in the left panel of Fig.~\ref{Fig:DJ2SchCh}),
which contradicts with the integrability of the Hamiltonian for the spinning
particle on the Schwarzschild background we prove in section~\ref{sec:RHamBL}.
It is true, however, that the relative error of the $J_{x},~J_{y}$ components,
and therefore of $J^2$ scale with $S$ (right panel of Fig.~\ref{Fig:DJ2SchCh}).
However, this should be anticipated since as $S\rightarrow 0$ the system 
basically ignores the spin contribution and tends to reproduce geodesic
trajectories.

In this work we focus on the properties of the equations of motion, and not so much
on the astrophysical implication of these equations. The spin of the particle
makes the trajectories to deviate from their geodesic paths. Thus, we can
interpret  the spin as a perturbation parameter of the system. A constant of
motion cannot dependent on the magnitude of the spin, even if the given value
might be astrophysically irrelevant. This independence from the spin magnitude
holds also for the integrability of a spinning particle Hamiltonian
(excluding of course the case when $S=0$). In our numerical calculations we
measure the spin in units of $m~M$, i.e. $\sigma=S/(m~M)$, and set both masses
to $1$, thus the spin parameter $\sigma=S$ is dimensionless. Large values of the
dimensionless spin, like $S=1$, might be astrophysically questionable, but do
not have implications on the dynamics (see, e.g., \cite{Suzuki97,Hartl03a} for
relevant discussions). In our paper large values of the spin serve mainly as
a tool to amplify the effects we want to point out, since these effects become 
less prominent when $S\ll 1$.

Moreover, we find that 
\[
\left\{ L_{x}^{2}+L_{y}^{2}+L_{z}^{2},H\right\} _{DB}=\mathcal{O}\left(S^{2}\right)~~,
\]
with 
\begin{align*}
L_{x} & =-p_{\theta}\sin(\phi)-p_{\phi}\cot(\theta)\cos(\phi)~~,\\
L_{y} & =p_{\theta}\cos(\phi)-p_{\phi}\cot(\theta)\sin(\phi)~~,\\
L_{z} & =p_{\phi}~~,
\end{align*}
which also have to be rewritten in terms of the canonical momenta $P_{i}$. The
conservation of the measure of the orbital momentum $L^{2}$ of the linearized in
spin MP equation in the case of the Schwarzschild spacetime background has been
thoroughly discussed in \cite{Apostolatos96} for the Pirani SSC \cite{Pirani56}.
When the measure of the spin $S^{2}=S_{I}S^{I}$ and the total angular momentum
$J^{2}$ are preserved, the integral of motion $L^{2}$ is equivalent to the
conservation of $\vec{L}\cdot\vec{S}$. However, as in the case of the total
angular momentum, we recover the same numerical problems for the measure of the
orbital angular momentum. These two oscillations issues can be traced back to
the coordinate dependence of the basis vectors in the spherical coordinate
system, as we will see in the next subsection.

So far, these coordinate effects have been investigated in Schwarzschild
spacetime. Since the Schwarzschild spacetime is the nonrotating limit
of the Kerr spacetime we would like to ensure that such coordinate
effects can be eliminated in the nonrotating limit, i.e., the coordinate
effects should vanish for nonrotating or slowly rotating black holes.
Thus, we were wondering whether there are more suitable choices of
a coordinate system and of a tetrad for rotating black holes which
do not show any unphysical coordinate effects in the Schwarzschild
limit. Hence, the question arises as to which coordinates are best
used?

Therefore, in the rest of Sec. \ref{sec:HamBL} we study the Hamiltonian
formulation in an isotropic coordinate systems for the same kind of
observer (ZAMO), introduced by \cite{Barausse10}.

\subsection{The Hamiltonian function in isotropic Cartesian coordinates}

\label{sec:RHamBL}

A revised Hamiltonian function for the Kerr spacetime background in
BL has been provided in \cite{Barausse10}. The formulation starts
in Cartesian quasi-isotropic coordinates. The line element in these
coordinates for an axisymmetric stationary metric is 
\begin{eqnarray}
ds^{2} & = & g_{tt}~dt^{2}\nonumber \\
 & + & 2~g_{tX}dX~dt+2~g_{tY}dY~dt+2~g_{XY}dX~dY\nonumber \\
 & + & g_{XX}~dX^{2}+g_{YY}~dY^{2}+g_{ZZ}~dZ^{2}~~,
\end{eqnarray}
with 
\begin{eqnarray}
g_{tt} & = & e^{-2\nu}\left[B^{2}\omega^{2}(X^{2}+Y^{2})-e^{4v}\right]~~,\nonumber \\
g_{tX} & = & e^{-2\nu}\omega B^{2}Y~~,\nonumber \\
g_{tY} & = & -e^{-2\nu}\omega B^{2}X~~,\nonumber \\
g_{XY} & = & -\frac{(e^{-2\nu}B^{2}-e^{2\mu})XY}{X^{2}+Y^{2}}~~,\nonumber \\
g_{XX} & = & \frac{e^{2\mu}X^{2}+e^{-2\nu}B^{2}~Y^{2}}{X^{2}+Y^{2}}~~,\nonumber \\
g_{YY} & = & \frac{e^{2\mu}Y^{2}+e^{-2\nu}B^{2}~X^{2}}{X^{2}+Y^{2}}~~,\nonumber \\
g_{ZZ} & = & e^{2\mu}~~
\end{eqnarray}
where $\omega,\, e^{\mu},\, e^{\nu},\, B$ are functions of $X,\, Y,\, Z$.

For this coordinate system the authors propose the tetrad field 
\begin{eqnarray}
\tilde{e}_{\beta}\,^{T} & = & e^{\nu}\delta_{\beta}^{t}~~,\nonumber \\
\tilde{e}_{\beta}\,^{1} & = & \frac{B~\omega~Y}{e^{\nu}}\delta_{\beta}^{t}+\frac{e^{\mu}X^{2}+e^{-\nu}B~Y^{2}}{X^{2}+Y^{2}}\delta_{\beta}^{X}\nonumber \\
 & + & \frac{(e^{\mu}-e^{-\nu}B)XY}{X^{2}+Y^{2}}\delta_{\beta}^{Y}~~,\nonumber \\
\tilde{e}_{\beta}\,^{2} & = & -\frac{B~\omega~X}{e^{\nu}}\delta_{\beta}^{t}+\frac{(e^{\mu}-e^{-\nu}B)XY}{X^{2}+Y^{2}}\delta_{\beta}^{X}\nonumber \\
 & + & \frac{e^{\mu}Y^{2}+e^{-\nu}B~X^{2}}{X^{2}+Y^{2}}\delta_{\beta}^{Y}~~,\nonumber \\
\tilde{e}_{\beta}\,^{3} & = & e^{\mu}\delta_{\beta}^{Z}~~,\label{eq:isoTetrad}
\end{eqnarray}
corresponding to an infalling observer with zero 3-momentum. This
tetrad becomes Cartesian, i.e., $\tilde{e}_{\beta}\,^{T}=1$, $\tilde{e}_{\beta}\,^{I}=\delta_{\beta}^{I}$,
in the flat spacetime limit.

The Cartesian quasi-isotropic coordinates relate with the BL coordinate
system through the transformation 
\begin{eqnarray}
X & = & R\left(r\right)\sin{\theta}\cos{\phi}~~,\nonumber \\
Y & = & R\left(r\right)\sin{\theta}\sin{\phi}~~,\nonumber \\
Z & = & R\left(r\right)\cos{\theta}~~,\nonumber \\
R\left(r\right) & = & \frac{1}{2}(r-M+\sqrt{\Delta})~~.\label{eq:coordTrafo}
\end{eqnarray}
The above relation between $r$ and $R$ holds outside the black hole's
horizon %
\footnote{The general relation between $r$ and $R$ is \\
 ${\displaystyle r=R+M+\frac{M^{2}-a^{2}}{4~R}}$ %
}.

% The parameters of Eq.~\eqref{eq:QILinEl} in BL are 
% \begin{eqnarray}
% B & = & \frac{\sqrt{\Delta}}{R}~~,\nonumber \\
% \omega & = & \frac{2aMr}{\Lambda}~~,\nonumber \\
% e^{2\nu} & = & \frac{\Delta\Sigma}{\Lambda}~~,\nonumber \\
% e^{2\mu} & = & \frac{\Sigma}{R^{2}}~~.
% \end{eqnarray}

When we go back to the Schwarzschild spacetime where $a\rightarrow0$,
\begin{equation}
ds^{2}=-f(R)dt^{2}+h(R)(dX^{2}+dY^{2}+dZ^{2})~~,\label{eq:isotropicSchw}
\end{equation}
the tetrad \eqref{eq:isoTetrad} reduces to the isotropic tetrad given
in \cite{Barausse09}: 
\begin{eqnarray}
\tilde{e}_{\beta}\,^{T} & = & \sqrt{1-\frac{2M}{r}}\delta_{\beta}^{t}=\sqrt{f\left(R\right)}\delta_{\beta}^{t}~~,\nonumber \\
\tilde{e}_{\beta}\,^{1} & = & \frac{r}{R}\delta_{\beta}^{X}=\left(1+\frac{M}{2R}\right)^{2}\delta_{\beta}^{X}~~,\nonumber \\
\tilde{e}_{\beta}\,^{2} & = & \frac{r}{R}\delta_{\beta}^{Y}=\left(1+\frac{M}{2R}\right)^{2}\delta_{\beta}^{Y}~~,\nonumber \\
\tilde{e}_{\beta}\,^{3} & = & \frac{r}{R}\delta_{\beta}^{Z}=\left(1+\frac{M}{2R}\right)^{2}\delta_{\beta}^{Z}~~,\label{eq:SchwLimitIsotropicTetrad}
\end{eqnarray}
where 
\begin{align*}
r & =R\left(1+\frac{M}{2R}\right)^{2}~~,\\
f(R) & =\frac{\left(2R-1\right)^{2}}{\left(2R+1\right)^{2}}~~,\\
h\left(R\right) & =\left(1+\frac{M}{2R}\right)^{4}~~.
\end{align*}
In order to check the behavior of these so called isotropic Cartesian
coordinates $\left(X,Y,Z\right)$ we analyze the conservation of the constants
of motion given by the symmetries of the system. The spherical symmetry
of the spacetime can be described in Cartesian-like coordinates $x^{\mu}$
by the following three Killing vectors 
\begin{equation}
{\xi}_{K}^{\mu}=\epsilon^{KLM}x^{L}\delta_{M}^{\mu}~~.\label{eq:xiXYZ}
\end{equation}
Using \eqref{eq:conservedC} we thus get the three conserved components
of the total angular momentum as a combination of the kinematical momentum
$p_{\mu}$ and the components of the spin tensor $S^{\mu\nu}$. On the other
hand, in the canonical description, the conservation of the total
angular momentum 
\begin{equation}
J_{K}=\epsilon^{KLM}x^{L}P_{M}+S_{K}~~,\label{eq:JschwXYZ}
\end{equation}
is demonstrated by vanishing Dirac brackets 
\begin{equation}
\left\{ J_{K},H\right\} _{DB}=0~~.
\end{equation}
Contrary to the previous case the canonical momenta $P_{M}$ and tetrad
components of the spin appear in this formula. The relations between
the two sets of quantities, the kinematical and the canonical, are
given by \eqref{eq:momentaHL} and \eqref{eq:SpinProj}. By computing
the difference of projection of \eqref{eq:conservedC} and \eqref{eq:JschwXYZ}
it can be shown, that if the Lie derivatives of the three spatial
tetrad vectors obey the Cartesian-like rule 
\begin{equation}
\left(\pounds_{\xi_{K}}\tilde{e}^{\mu}\,_{L}\right)\tilde{e}_{\mu M}=-\epsilon_{KLM}~~~~{\rm and}~~~~\xi_{K}^{0}\equiv0~~,\label{eq:tetradSphericalSymComptibile}
\end{equation}
the two conserved quantities, one in kinematical variables and the
other in canonical ones, are identical. Indeed, this formula holds
in the flat Minkowski spacetime for the Cartesian tetrad 
$\tilde{e}_{\mu}\,^{A}=\delta_{\mu}^{A}$, which naturally leads to the intuition,
that a tetrad, that reduces to a Cartesian one in the flat spacetime, is a good 
tetrad choice. (In \eqref{eq:tetradSphericalSymComptibile} the fact that the 
time component of the Killing vectors is required to vanish is explicitly stated,
since it is written as a covariant, coordinate independent formula,
but it has been derived using the above coordinate assumption.)

The general condition \eqref{eq:tetradSphericalSymComptibile} can
now be applied to the particular case of the Schwarzschild limit \eqref{eq:isotropicSchw}.
As Lie derivatives can be written using partial rather than covariant
derivatives, one can easily check, that the tetrad field \eqref{eq:SchwLimitIsotropicTetrad}
satisfies \eqref{eq:tetradSphericalSymComptibile}.

Yet, as an example, that the equivalence between total angular momentum
expressed in kinematical and canonical variables is not so obvious,
let us consider a symmetry of the Schwarzschild spacetime w.r.t. rotation
along the $z$-axis 
\begin{equation}
{\xi}_{z}^{\mu}=[0,-Y,X,0]~~.
\end{equation}
It yields the related component of the total angular momentum 
\begin{align*}
J_{z}= & Xp_{y}-Yp_{x}+S^{xy}\left(h\left(R\right)+\frac{\left(h\left(R\right)\right)'}{2R}\left(X^{2}+Y^{2}\right)\right)\\
 & -\frac{\left(h\left(R\right)\right)'}{2R}Z\left(XS^{yz}+YS^{zx}\right)~~.
\end{align*}
Here, $p_{i}$ represents the kinematical MP momenta and $S^{ij}$
the coordinate spin components, the prime denotes a derivative with
respect to $R$. In the Hamiltonian approach we use the canonical
momenta $P_{i}$ and the projected spin components $S^{I}$, so it
is necessary to perform a transformation from $\left(p_{i},S^{ij}\right)$
to $\left(P_{I},S^{I}\right)$ using the relations given in \eqref{eq:momentaHL}
and \eqref{eq:SpinProj}. By doing this, terms proportional to $h'(R)$
get absorbed into $P_{x}$ and $P_{y}$ and the corresponding component
of the total angular momentum can be written as 
\begin{align}
J_{z} & =XP_{y}-YP_{x}+S_{3}~~.\label{eq:JzschwXYZ}
\end{align}
The corresponding Hamiltonian in these coordinates, cf.~\cite{Barausse09}
reads 
\[
{H=H_{NS}+H_{S}~~,}
\]
with 
\begin{align}
H_{NS} & =\frac{1}{\sqrt{f\left(R\right)}}\sqrt{Q}~~,\label{eq:HNSisotropic}\\
H_{S} & =\frac{1-\frac{M}{2R}+2\left(1-\frac{M}{4R}\right)\sqrt{Q}}{\left(1+\frac{M}{2R}\right)^{6}R^{3}\sqrt{Q}\left(1+\sqrt{Q}\right)}\frac{M}{m}\left(\vec{L}\cdot\vec{S}\right)~~,\label{eq:HSisotropic}
\end{align}
and $Q=m^{2}+\frac{1}{h\left(R\right)}\vec{P}^{2}$. Notice, that
setting $M\rightarrow0$, i.e., no gravitational field, we indeed
obtain that the spin part of the Hamiltonian $H_{S}$ becomes zero,
as it should in the Minkowskian spacetime.

Next, we can easily compute the evolution equations for the $J_{x}$,
$J_{y}$ and $J_{z}$ as 
\begin{align*}
\left\{ J_{x},H\right\} _{DB} & =0~~,\\
\left\{ J_{y},H\right\} _{DB} & =0~~,\\
\left\{ J_{z},H\right\} _{DB} & =0~~,
\end{align*}
which is thus also true for the measure of the total angular momentum
$J^{2}$. Moreover, it holds that $\left\{ L^{2},H\right\} =0$ where
$L^{2}=L_{x}^{2}+L_{y}^{2}+L_{z}^{2}$ is the measure of the orbital
angular momentum. Its respective components are defined as $L_{i}=\varepsilon_{ijk}q^{j}P^{k}$,
with $q^{i}=\left(X,Y,Z\right)$ and $P^{i}=\left(P_{x},P_{y},P_{z}\right)$.

\begin{figure}[htp]
\centerline{ \includegraphics[width=0.45\textwidth]{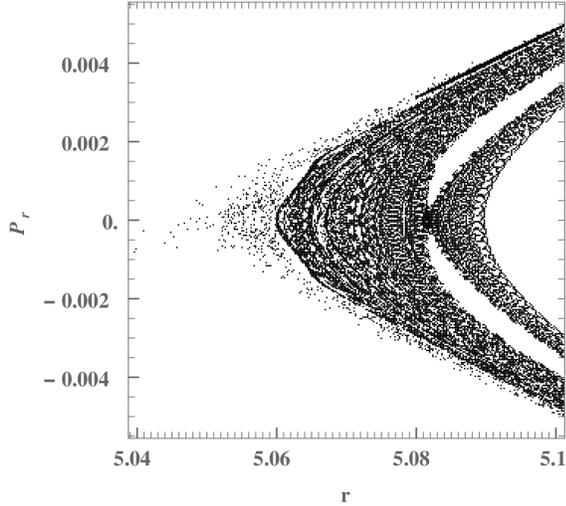}}
\caption{A detail from the surface of section $\theta=\pi/2$, $P_{\theta}>0$.
The parameters of the orbits are $H=0.9449111825230683$, $J_{z}=3.5$,
$S=M=m=1$, $a=0.1$, the common initial conditions are $\phi=0$,
$P_{r}=0$, $S_{1}=0$, while by solving numerically the system $P_{\theta}=-S_{2}$,
$J_{z}=P_{\phi}+S_{3}$, and $S=\sqrt{S_{2}^{2}+S_{3}^{2}}$ we define
the rest. }

\label{Fig:KerrChaos} 
\end{figure}

In fact, since the linearized in spin Hamiltonian system  given by
\eqref{eq:HNSisotropic}, \eqref{eq:HSisotropic} has five degrees of freedom, 
the five independent and in involution constants of motion
$(J_{z},J^{2},L^{2},S^{2},H)$ of the Schwarzschild limit make the system
integrable. The integrability for the Schwarzschild background seems to result
from the linearized in spin Hamiltonian approximation, because in~\cite{Suzuki97}
it has been shown that for the full MP equations with T SSC chaos appears for a
spinning particle moving in the Schwarzschild background. However, the
integrability seems to vanish in the Hamiltonian approximation once we turn on
the spin of the central body. Namely, in the case of a Kerr spacetime
background chaos appears again (scattered dots in Fig.~\ref{Fig:KerrChaos}),
which suggests the nonintegrability of the corresponding Hamiltonian
system. The existence of chaos in the Kerr background case for the
Hamiltonian approximation is not just a confirmation of previous studies
concerning the full MP equations with T SSC, see, e.g., \cite{Hartl03a,Hartl03b}.
It shows that the linearized in spin Hamiltonian function given in
\cite{Barausse10} is non-integrable. This result contradicts statements in the
literature saying that up to the linear order in spin the motion of a spinning
particle corresponds to an integrable system, see, e.g., \cite{Hinderer13}.
A thorough investigation of chaos in the Kerr spacetime for the linearized in
spin Hamiltonian function will be provided in \cite{LGKPS}.

The above results match exactly the expectations we had from the symmetries.

Up to this point, we have discussed the properties of a ZAMO tetrad in spherical
and Cartesian coordinates in Schwarzschild spacetime. Taking the conservation
of the constants of motion for numerical calculations as an important
criterion to be satisfied, promising indicators for a ``good'' tetrad
choice are the reduction to Cartesian tetrad in flat spacetime as
well as the vanishing of the spin dependent Hamiltonian. Two questions
arise with this statement: First, are there other coordinates we may
choose providing us with ``good'' tetrads, and second, since we
were focusing on a ZAMO tetrad, we ask whether a non-ZAMO tetrad yields
the same properties if the coordinate basis is not changed. We expect
the properties of the tetrad to depend on the choice of the coordinates,
so that in the following we take Kerr-Schild coordinates and analyze
two tetrads, one ZAMO and one non-ZAMO tetrad.

\section{The Hamiltonian function in Kerr-Schild coordinates}

\label{sec:RHamKS}

The Kerr-Schild coordinates have the great advantage
that they are horizon penetrating so that they are well behaved in
the vicinity of the horizon, which simplifies numerical calculations
in this domain, probably improving the numerical treatment compared
to isotropic coordinates for events in the strong field. Here we shall
introduce a Hamiltonian function using the Kerr-Schild (KS) coordinates.
The line element in KS coordinates reads \cite{Gualtierie} 
\begin{eqnarray}
ds^{2} & = & g_{\mu\nu}d\bar{x}^{\mu}d\bar{x}^{\nu}~~,\nonumber \\
g_{\mu\nu} & = & \eta_{\mu\nu}+f~l_{\mu}~l_{\nu}~~,
\end{eqnarray}
where $(0,1,2,3)$ correspond to $(\bar{t},\bar{x},\bar{y},\bar{z})$,
\begin{eqnarray}
l_{\bar{t}} & = & -1~~,\nonumber \\
l_{\bar{x}} & = & -\frac{\bar{r}~\bar{x}+a~\bar{y}}{\bar{r}^{2}+a^{2}}~~,\nonumber \\
l_{\bar{y}} & = & -\frac{\bar{r}~\bar{y}-a~\bar{x}}{\bar{r}^{2}+a^{2}}~~,\nonumber \\
l_{\bar{z}} & = & -\frac{\bar{z}}{\bar{r}}~~,\label{eq:Deflx}\\
\end{eqnarray}
and 
\begin{eqnarray}
f & = & \frac{2~M~\bar{r}^{3}}{\bar{r}^{4}+a^{2}~\bar{z}^{2}}~~,\label{eq:Deff}\\
\bar{r} & = & \sqrt{\frac{\bar{\rho}^{2}+\sqrt{\bar{\rho}^{4}+4a^{2}~z^{2}}}{2}}~~,\nonumber \\
\bar{\rho}^{2} & = & \bar{x}^{2}+\bar{y}^{2}+\bar{z}^{2}-a^{2}~~.
\end{eqnarray}
For simplicity in the rest of the section we drop the bar notation
over the KS coordinates.

Independently on the tetrad field the choice of coordinates implies
the non-spinning part of the Hamiltonian 
\begin{equation}
H_{NS}=\alpha^2 f~l_{i}P_{i}+\alpha\sqrt{m^{2}+P_{i}P_{i}-f\alpha^{2}(l_{i}P_{i})^{2}}~~,\label{eq:HNSKS}
\end{equation}
where $l_{i}P_{i}=\delta^{ij}l_{i}P_{j}$ and 
\begin{equation}
\alpha=\frac{1}{\sqrt{1+f}}~~.
\end{equation}
In the following we present two tetrad choices corresponding to different
types of observers.

\subsection{ZAMO Tetrad}

\label{sec:ZamoTed} In the previous section, we focused on a tetrad
field associated to the observers with vanishing momenta $P_{i}=0$,
i.e., zero angular momentum observers (ZAMO), in two different coordinate
systems, isotropic Cartesian and Boyer-Lindquist coordinates. Therefore,
it is reasonable to first consider such an observer in KS coordinates
as well. Here, we choose a tetrad corresponding to an observer infalling
with the radial velocity $U^{r}=-\alpha f$: 
\begin{eqnarray*}
\tilde{e}_{\beta}\,^{T} & = & \alpha\;\delta_{t}^{0}~~,\\
\tilde{e}_{\beta}\,^{I} & = & \delta_{\beta}^{I}+\left(\alpha^{-1}\!\!-1-\alpha f\right)l_{I}\,\delta_{\beta}^{0}+\left(\alpha^{-1}\!\!-1\right)l_{I}l_{\beta}~~.
\end{eqnarray*}
Again, this tetrad becomes Cartesian, i.e., $\tilde{e}_{\beta}\,^{T}=\delta_{\beta}^{T}$,
$\tilde{e}_{\beta}\,^{I}=\delta_{\beta}^{I}$, in the flat spacetime
limit, which is a first indicator for being a good tetrad choice.
The next step is to analyze the behavior in the Schwarzschild limit
$a\rightarrow0$. Then, following the procedure introduced in \cite{Barausse09},
we obtain the Hamiltonian $\bar{H}^{{\rm Schw}}=\bar{H}_{NS}^{{\rm Schw}}+\bar{H}_{S}^{{\rm Schw}}$
with 
\begin{align}
{\bar{H}}_{NS}^{{\rm Schw}} & =\alpha\left(\overline{m}-\frac{2M\alpha}{r^{2}}\;\vec{r}\!\cdot\!\vec{P}\right)~~,\label{eq:HnsZAMO}\\
{\bar{H}}_{S}^{{\rm Schw}} & =\frac{M}{\overline{m}}\left[\frac{2\alpha^{2}}{\alpha+1}-\frac{\alpha^{5}+3\alpha^{3}}{r}\frac{\vec{r}\!\cdot\!\vec{P}}{\omega_{T}}-\alpha^{4}\frac{\overline{m}}{\omega_{T}}\right]\frac{\vec{L}\cdot\vec{S}}{r^{3}}~~,\label{eq:HsZAMO}
\end{align}
where 
\begin{align}
\overline{m} & =\sqrt{m^{2}+\vec{P}^{2}-\frac{f\alpha^{2}}{r^{2}}\left(\vec{r}\!\cdot\!\vec{P}\right)^{2}}~~,\nonumber \\
\omega_{T} & =-m-\overline{m}~~.\label{eq:barm}
\end{align}
Since the total Hamiltonian is merely a function of certain scalar
combinations of $\left(\vec{r},\vec{P},\vec{S}\right)$ (where $\vec{r}=\left(x,y,z\right)$),
namely $\bar{H}=\bar{H}\left(|\vec{r}|^{2},|\vec{P}|^{2},\vec{r}\cdot\vec{P},\vec{L}\cdot\vec{S}\right)$
with $L_{i}=\varepsilon_{ijk}r^{j}P^{k}$ , we can deduce that 
\begin{equation}
\left\{ \vec{L}+\vec{S},H\right\} _{DB}=0~~,\label{eq:LSHbracket}
\end{equation}
by using the canonical structure of the variables. Moreover, we would
like to stress here again, that the conservation of $L^{2}$ in Schwarzschild
spacetime is equivalent to the conservation of $\vec{L}\cdot\vec{S}$
so that it suffices to express the Hamiltonian in terms of $\vec{L}\cdot\vec{S}$
in order to show \eqref{eq:LSHbracket}. In fact, it reflects the
integrability of the system at linear order in spin.

However, we cannot simply infer that $\vec{J}=\vec{L}+\vec{S}$ is
valid in the new canonical coordinates. The conserved total angular
momentum is already given by \eqref{eq:conservedC} and by the Killing
vectors \eqref{eq:xiXYZ} of the Schwarzschild spacetime we get 
\begin{align}
J_{i} & =\tilde{L}_{i}+S_{i}~~,\label{eq:JSchwKS}
\end{align}
where the tilde denotes the quantities to be written in terms of the
kinematical momenta $p_{i}$ $\left(L_{i}=\varepsilon_{ijk}r^{j}p^{k}\right)$
and the index $i$ in $S_{i}$ refers to the coordinate basis. This
relation is valid in KS coordinates, independent of the tetrad choice.

\begin{figure}[htp]
\centerline{ \includegraphics[width=0.25\textwidth]{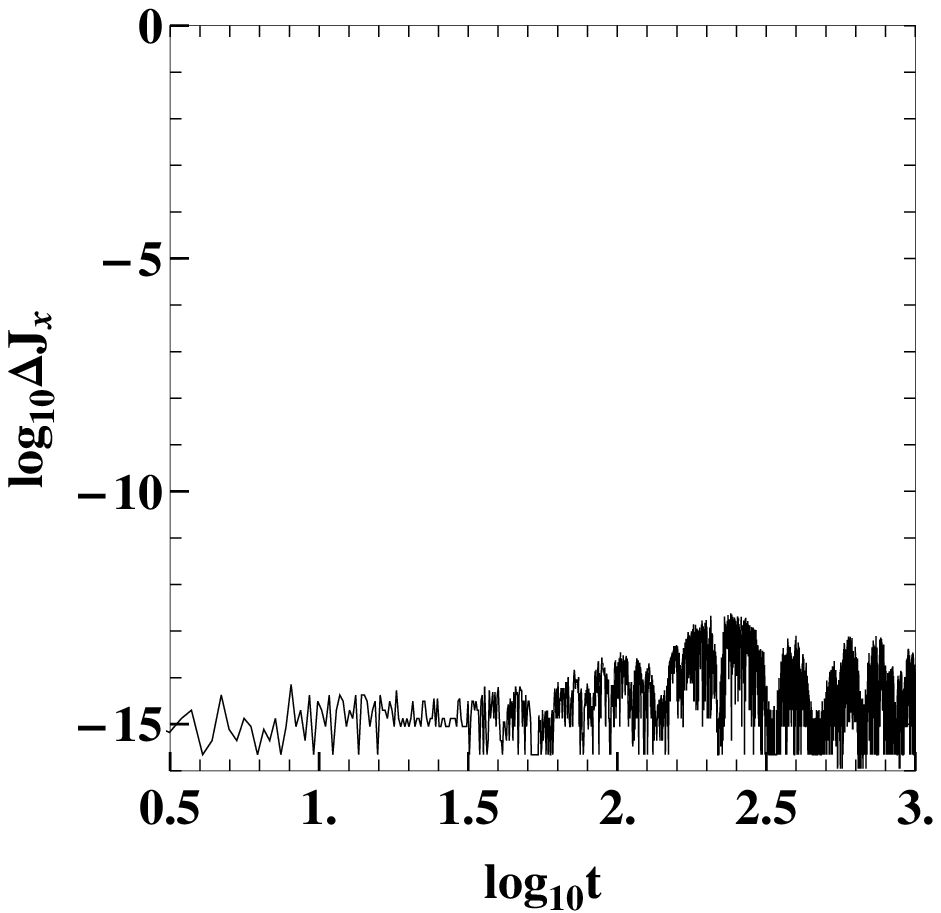} \includegraphics[width=0.25\textwidth]{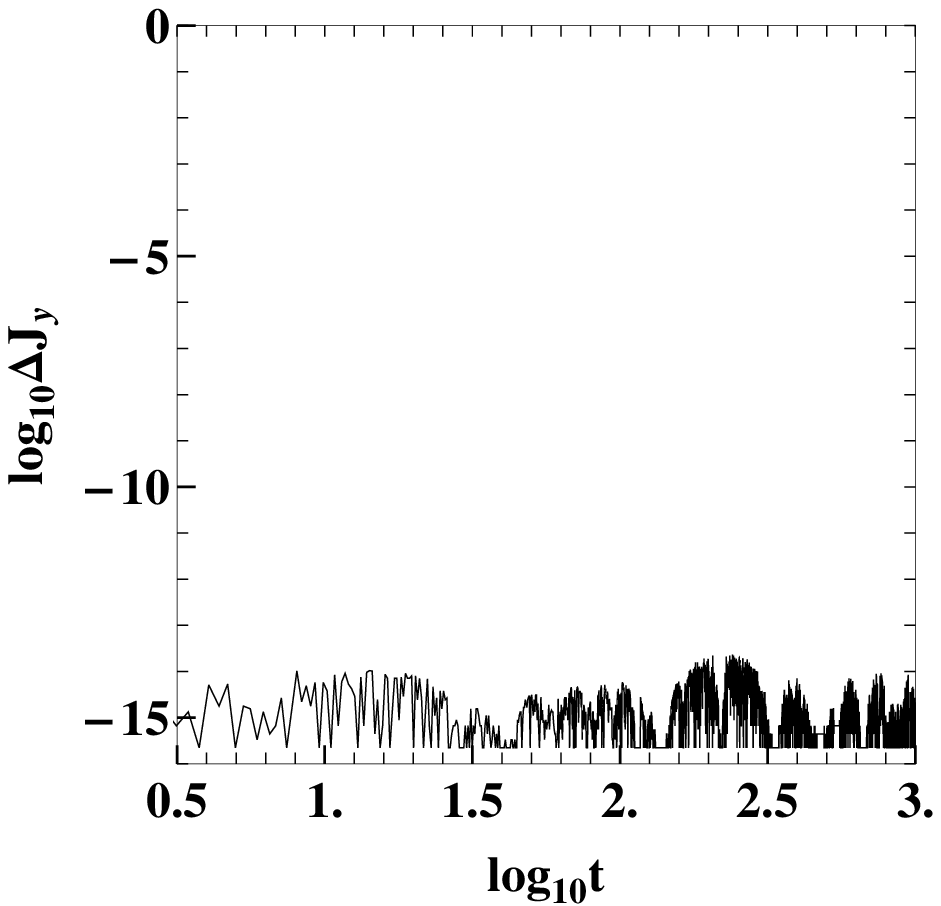}}
\caption{The left panel shows the relative error of $J_{x}$, and the right
of $J_{y}$ as a function of time in logarithmic scale for the ZAMO
tetrad in KS coordinates as evolved by the Hamiltonian with $a=0$,
$M=m=1$, and $S=1$.}

\label{Fig:DJxDJyKSZAMO} 
\end{figure}

In order to relate the conserved quantities to the canonical momenta
$P_{i}$ and the tetrad components of the spin $S_{I}$, we have to
perform a transformation from $\left(p_{i},S_{i}\right)$ to $\left(P_{i},S_{I}\right)$
using the relations given in \eqref{eq:momentaHL} and \eqref{eq:SpinProj}.
Therewith, we indeed find the components $J_{i}$ to be given by \eqref{eq:JzschwXYZ},%
\begin{comment}
\begin{align*}
J_{x} & =L_{x}+S_{1}~~,\\
J_{y} & =L_{y}+S_{2}~~,\\
J_{z} & =L_{z}+S_{3}~~,
\end{align*}
and the corresponding evolution equations can now be computed by 
\[
\left\{ J_{i},H\right\} _{DB}=0~~,
\]
\end{comment}
{} which yields vanishing Dirac brackets for each component of the total
angular momentum according to the argument mentioned in Section \ref{sec:RHamBL}.
In order to support this statement we performed a numerical check
shown in Fig.~\ref{Fig:DJxDJyKSZAMO}. 

It is immediately obvious that the conservation of these components
is ensured up to numerical errors which do not accumulate over the
integration time but stay at the same level. These results are similar
to the ones obtained in isotropic Cartesian coordinates, so that the
quality of the outputs is comparable. Therefore, if one can choose
between KS and isotropic Cartesian coordinates, there is no preferred
choice between those two in Schwarzschild spacetime. However, if the
dynamics of plunging orbits is considered in a Kerr spacetime background,
it may be more sensible to change to KS coordinates since they are
horizon penetrating and avoid numerical divergence close to the horizon
(see Appendix~\ref{sec:PlOr}).

Second, we consider the contribution from the spin part of the Hamiltonian
in flat spacetime. From \eqref{eq:HsZAMO} we see that for $M\rightarrow0$
the contributions from $H_{S}$ vanish as it should. Hence, also additional
coordinate effects which arise in spherical coordinates are avoided,
further supporting such a choice of tetrad and coordinates.

\subsection{Non-ZAMO tetrad}

\label{sec:NonZamo}

To simplify the Hamiltonian in Kerr-Schild coordinates we change to
another tetrad field, which is not required to be a ZAMO observer.
In particular, we take advantage of the fact that for certain observers
no square roots appear due to normalization of the tetrad vectors
\begin{align}
\tilde{e}_{\mu}\,^{T} & =\left[1-\frac{f}{2},~\frac{f}{2}l_{x},~\frac{f}{2}l_{y},~\frac{f}{2}l_{z}\right]~~,\\
\tilde{e}_{\mu}\,^{X} & =\left[~-\frac{f}{2}l_{x},~1+\frac{f}{2}l_{x}l_{x},~\frac{f}{2}l_{x}l_{y},~\frac{f}{2}l_{x}l_{z}\right]~~,\\
\tilde{e}_{\mu}\,^{Y} & =\left[~-\frac{f}{2}l_{y},~\frac{f}{2}l_{y}l_{x},~1+\frac{f}{2}l_{y}l_{y},~\frac{f}{2}l_{y}l_{z}\right]~~,\\
\tilde{e}_{\mu}\,^{Z} & =\left[~-\frac{f}{2}l_{z},~\frac{f}{2}l_{z}l_{x},~\frac{f}{2}l_{z}l_{y},~1+\frac{f}{2}l_{z}l_{z}\right]~~,
\end{align}
where we use the definitions from above, cf.~Eqs.~\eqref{eq:Deflx}-\eqref{eq:Deff}.
This is the tetrad of an infalling `non-ZAMO' observer, as the observer's
specific angular momentum 
\begin{align}
\tilde{e}_{\phi T}=\left(\frac{\partial}{\partial\phi}\right)^{\mu}\tilde{e}_{\mu T}=-\frac{1}{2}\,\frac{fa}{r}\frac{x^{2}+y^{2}}{r^{2}+a^{2}} & \ne0~~,
\end{align}
and the observer's radial coordinate velocity 
\[
\tilde{e}^{r}\,_{T}=\left(\frac{\partial r}{\partial x^{\mu}}\right)\tilde{e}^{\mu}\,_{T}~=-\frac{f}{2}<0~~.
\]
Thus, we again compute the Hamiltonian in canonical coordinates up
to linear order in spin given by \cite{Barausse09} 
\begin{align}
H=H_{NS}+H_{SO}+H_{SS}~~,
\end{align}
where $H_{NS}$ is given by \eqref{eq:HNSKS}, 
\begin{align}
H_{SO}= & ~\alpha f\frac{Mm-2{\widetilde{m}}(M-fr)}{2M\;\overline{m}\;\omega_{T}}~\frac{r\;\epsilon^{ijK}l_{i}p_{j}S^{K}}{r^{2}+a^{2}l_{z}^{2}}~~,
\end{align}
and 
\begin{align}
 & H_{SS}=~-\frac{af}{4\,\omega_{T}~M{\overline{m}}~(a^{2}l_{z}^{2}+r^{2})}\times\Bigg\{\\
 & \Big[4\,{{f}{l_{z}}\,{\widetilde{m}}\left(({\overline{m}}-\alpha f{\widetilde{m}}){r}+\alpha{\widetilde{m}}M\right)}-2M\,{l_{z}}\left({\overline{m}}m+\alpha{m}^{2}\right)\nonumber \\
 & ~+2\,{\alpha\left[(M+{2\, r})mf{l_{z}}+(3M-2{fr})P_{z}\right]{\widetilde{m}}}\Big]S^{i}l_{i}\nonumber \\
 & +2\,\alpha\left(m+{\widetilde{m}}\right)\left[M{l_{z}}\;{S^{i}P_{i}}-{\widetilde{m}}\left(2\,{fr}-3M\right){S^{3}}\right]\nonumber \\
 & -2\,{\frac{{a}{l_{z}}\,{\widetilde{m}}}{{r}^{2}}}\left(3\, Mr-{a}^{2}f{l_{z}}^{2}-3\, f{r}^{2}\right)\Big[\alpha(S^{1}P_{y}-S^{2}P_{x})\nonumber \\
 & -\left(\alpha m+{\overline{m}}-\alpha f{\widetilde{m}}\right)(S^{1}l_{y}-S^{2}l_{x})\Big]\Bigg\}~~.\nonumber 
\end{align}
Here, instead of \eqref{eq:barm}, we used 
\begin{align}
\overline{m} & =\sqrt{m^{2}+P_{i}P_{i}-f\alpha^{2}(l_{i}P_{i})^{2}}~~,\nonumber \\
{\widetilde{m}} & =\alpha\overline{m}-\alpha^{2}P_{i}l^{i}~~,\nonumber \\
\omega_{T} & =-m-\frac{\overline{m}}{\alpha}+\frac{f}{2}\widetilde{m}~~,\label{eq:omegaT}
\end{align}
which together with the usage of components of $l^{\mu}$ instead
of coordinates significantly shortened expressions for $H_{SO}$ and
$H_{SS}$. All vector components are grouped in such a way that the
relation $\left\{ L_{z}+S_{z},H\right\} _{DB}=0$ is obvious.

\begin{figure}[htp]
\centerline{ \includegraphics[width=0.25\textwidth]{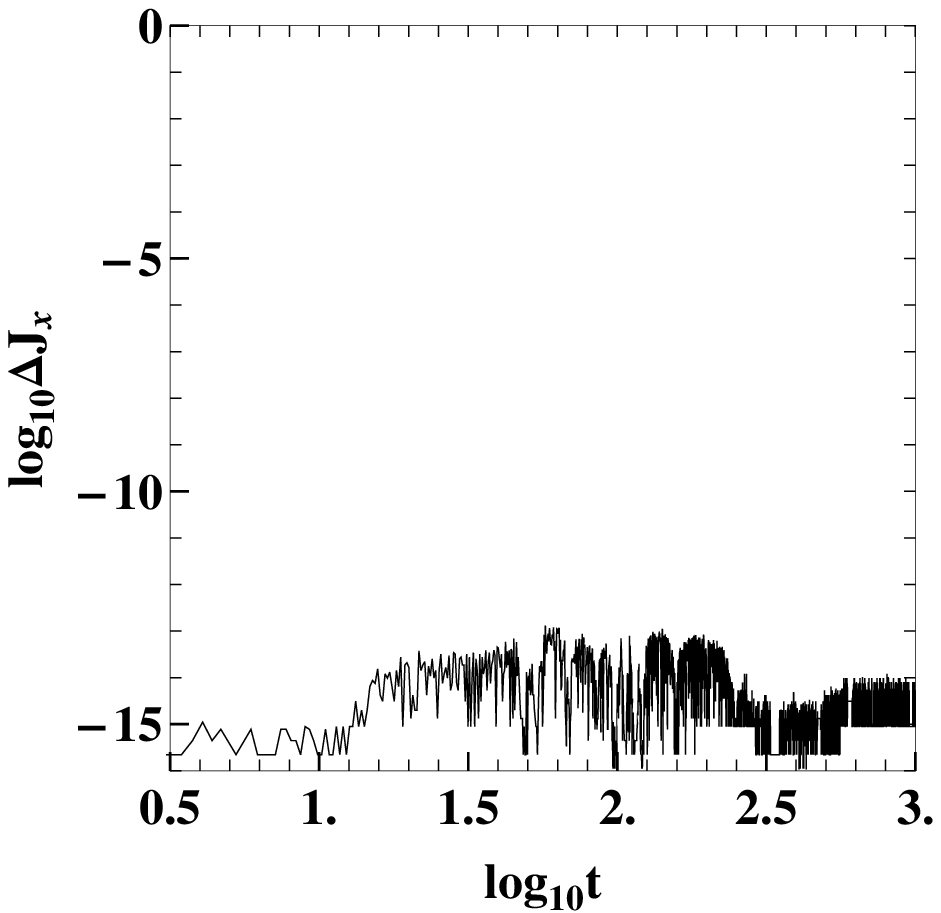}
\includegraphics[width=0.25\textwidth]{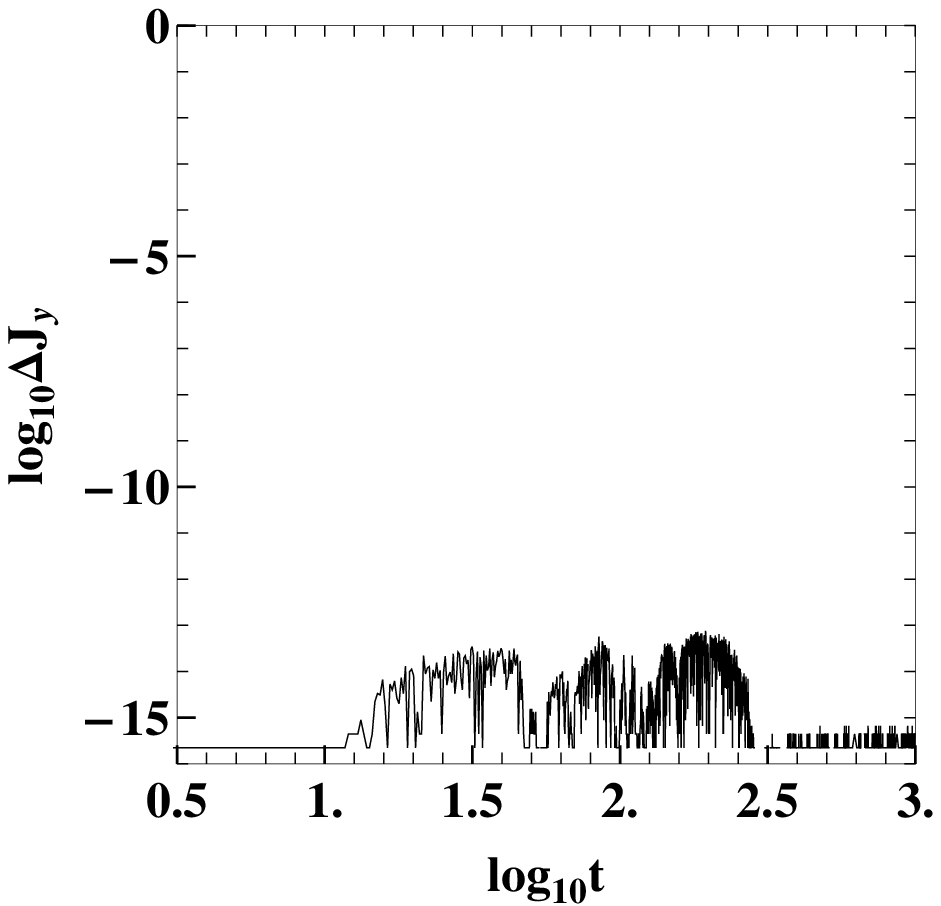}}
\caption{The left panel shows the relative error of $J_{x}$, and the right
of $J_{y}$ as a function of time in logarithmic scale for the non-ZAMO
tetrad in KS coordinates as evolved by the Hamiltonian with $a=0$,
$M=m=1$, and $S=1$.}

\label{Fig:DJxDJyNonZAMOKS} 
\end{figure}

Again, the complete angular momentum conservation is restored in the
Schwarzschild limit. Since $\bar{H}_{NS}$ only depends on the chosen
coordinate basis, it is still given by \eqref{eq:HnsZAMO}. The spinning
part 
\begin{align}
\bar{H}_{S} & =\left[\alpha\frac{M}{\overline{m}}\left(1\!-\!\frac{M+2r}{r\left(r+2M\right)}\frac{\vec{r}\!\cdot\!\vec{P}}{\omega_{T}}\right)\!-\!\frac{M}{\omega_{T}}\frac{1-\frac{M}{r}}{1+\frac{2M}{r}}\right]\frac{\vec{L}\cdot\vec{S}}{r^{3}}~~,\label{eq:HsNonZamo}
\end{align}
where $\bar{m}$ and $\omega_{T}$ are given by \eqref{eq:omegaT},
can again be written as a function $\bar{H}=\bar{H}\left(|\vec{r}|^{2},|\vec{P}|^{2},\vec{r}\cdot\vec{P},\vec{L}\cdot\vec{S}\right)$
so that we can follow the reasoning of the preceding subsection to
obtain vanishing Dirac brackets \eqref{eq:LSHbracket}. Therefore,
we only have to check the equations for the components of the total
angular momentum $J_{i}$ in canonical coordinates $\left(P_{i},S_{I}\right)$.
Using the expressions for the total angular momentum with respect
to the coordinate basis \eqref{eq:JSchwKS}, we again perform a transformation
to the tetrad basis and the canonical momenta and recover relation
\eqref{eq:JzschwXYZ}. Thus, in the Schwarzschild limit, the non-ZAMO
tetrad in KS coordinates has the same numerical properties as the
ZAMO tetrad, as expected, which is also visible in Fig.~\ref{Fig:DJxDJyNonZAMOKS}.

Consequently, it seems to be a good choice of coordinate system for
numerical investigations.

It is of course also possible to rewrite the coefficients of the tetrad
basis vectors in terms of any coordinate system without changing the
general properties of the Hamiltonian system as long as the tetrad
basis vectors remain oriented along the isotropic coordinate (Cartesian
like) basis vectors. In \cite{Barausse09}, it was already mentioned
that the coordinate effects can be avoided by choosing the directions
of the tetrad basis vectors along a Cartesian coordinate system. However,
if the tetrad corresponds to a Cartesian frame, the spin variables
remain Cartesian whereas the position and momentum variables are spherical
ones. This approach is used in effective-one-body theory or post-Newtonian
methods in order to compare the dynamical contributions from different
orders in spin, (see e.g., \cite{Porto06,Barausse09}) and may in
fact also be used for the computation of the equations of motion from
the Hamiltonian. Nevertheless, in that case it is more sensible to
be consistent in the choice of coordinates and spin variables so that
the Dirac brackets can be used for the calculation of the equations
of motion. This coordinate system does not necessarily adapt to the
symmetries of the spacetime as we have seen. Generally, it is very
useful to choose a coordinate system and corresponding basis vectors
that do not imply coordinate effects if one aims at the analysis of
the equations of motion.

\section{Conclusions}

\label{sec:Conclusions}

In this work we have studied the Hamiltonian formalism of a spinning
particle provided in \cite{Barausse09} with regard to numerical investigations
of the equations of motion. It was already discussed in \cite{Barausse09}
that this Hamiltonian formalism does not only depend on the tetrad
field one uses, but also on the coordinate system one chooses in order
to express the tetrad field or the Hamiltonian function. Using the
Dirac brackets to check the integrals of motion, we have shown that
an unfortunate choice of the coordinate system can lead to a nonpreservation
of quantities in numerical integration which should, according to
the symmetries of the system and the linearized MP equations, be conserved.
However, we find that the type of the tetrads, i.e., whether the observer
is ZAMO or follows some other worldline which does not correspond
to a ZAMO does not affect the general dynamical properties of the
constants of motion. In fact, we have examined both kinds of tetrads
and found no difference in their ability to be numerically applied,
i.e., they possess the same properties with respect to numerical computations.
However, the formulae for the spinning part of the Hamiltonian can
be simplified and compactified, which we think is worthwhile to be mentioned.

In order to obtain Hamiltonian systems without coordinate effects
smearing the actual physical behavior in numerical solutions and still
reliable in the vicinity of the central object's horizon, two new
horizon penetrating Hamiltonian functions were introduced. Both of
them were constructed on tetrad fields which were expressed in Kerr-Schild
coordinates. In particular, the non-ZAMO tetrad allows us to express
the Hamiltonian both in Schwarzschild and Kerr spacetime in a simple
and compact form. Future (numerical) work may profit from this explicit
Hamiltonian.

While studying the Dirac brackets in the Schwarzschild limit, we have
shown that in this limit the Hamiltonian functions with acceptable
properties are integrable. In particular, we have shown that the system's
five degrees of freedom admit five independent and in involution
integrals of motion. On the other hand, we have shown by a numerical
example that chaos appears when we go to the Kerr background in the
case of the Hamiltonian function proposed in \cite{Barausse10}. This
suggests that this linearized Hamiltonian function corresponds to
a nonintegrable system in the case of the Kerr background. Indeed,
the Hamiltonian formulation offers a wide range of applications in
the context of chaos and perturbation theory, such as Poincaré sections
or recurrence plots. In order to answer the question for chaos thoroughly,
a detailed analysis of the motion of spinning particles in the Kerr
spacetime described by the linearized in spin Hamiltonian approximation
is in progress \cite{LGKPS}.
\begin{acknowledgments}
G.L-G and J.S. were supported by the DFG grant SFB/Transregio 7. G.L-G
is supported byUNCE-204020. T.L. and G.L-G were supported by GACR-14-10625S.
D.K. gratefully acknowledges the support from the Deutsche Forschungsgemeinschaft
within the Research Training Group 1620 ``Models of Gravity'' and
from the ``Centre for Quantum Engineering and Space-Time Research
(QUEST)''. 
\end{acknowledgments}

\appendix
%dummy comment inserted by tex2lyx to ensure that this paragraph is not empty
%dummy comment inserted by tex2lyx to ensure that this paragraph is not empty
%dummy comment inserted by tex2lyx to ensure that this paragraph is not empty
%dummy comment inserted by tex2lyx to ensure that this paragraph is not empty
%dummy comment inserted by tex2lyx to ensure that this paragraph is not empty

\section{Numerical integration of the Hamiltonian equations}

\label{sec:NumMet} 

Our numerical integrators rely on the considerations
of our previous work, cf.~\cite[Secs.~A~and~B]{LSK1}, which can
be extended to all Hamiltonians in this work. Let us briefly summarize
the main points.

All Hamiltonian equations of this work possess a so-called \textit{Poisson
structure}, i.e., for $\mathbf{y}=(P_{1},P_{2},P_{3},x^{1},x^{2},x^{3},S_{1},S_{2},S_{3})^{T}\in\mathbb{R}^{9}$,
they can be written as 
\begin{align}
\dot{\mathbf{y}}=B(\mathbf{y})\grad H(\mathbf{y})~~,\label{def-poisson-sys}
\end{align}
where $B:\mathbb{R}^{9}\to\mathbb{R}^{9\times9}$ is the skew-symmetric
matrix-valued function 
\begin{align}
B(\mathbf{y}) & =\begin{pmatrix}0 & -I_{3\times3} & 0\\
I_{3\times3} & 0 & 0\\
0 & 0 & B_{1}(\mathbf{y})
\end{pmatrix}~~,
\end{align}
with 
\begin{align}
I_{3\times3} & =\begin{pmatrix}1 & 0 & 0\\
0 & 1 & 0\\
0 & 0 & 1
\end{pmatrix}~~,\\
B_{1}(\mathbf{y}) & =\begin{pmatrix}0 & -S_{3} & S_{2}\\
S_{3} & 0 & -S_{1}\\
-S_{2} & S_{1} & 0
\end{pmatrix}~~.
\end{align}
Due to this special structure, the spin length $S=\sqrt{S_{1}^{2}+S_{2}^{2}+S_{3}^{2}}$
is conserved along solutions of the equations of motion~\eqref{def-poisson-sys}.
Thus, the three dimensional spin $\mathbf{S}=(S_{1},S_{2},S_{3})^{T}$
can be represented by two variables $\alpha$ and $\xi$ via 
\begin{align}
\mathbf{S}=S\begin{pmatrix}\sqrt{1-\xi^{2}}\cos(\alpha)\\
\sqrt{1-\xi^{2}}\sin(\alpha)\\
\xi
\end{pmatrix}~~,
\end{align}
see,~e.g.,~\cite{wuxie,Seyrich13}. One can then show, cf.~\cite{Seyrich13},
that 
\begin{align}
\dot{\xi} & =-\pd H\alpha~~,\\
\dot{\alpha} & =\pd H\xi~~,
\end{align}
holds. Hence, in the transformed variables $\mathbf{z}=(P_{1},P_{2},P_{3},\xi,x^{1},x^{2},x^{3},\alpha)$,
the equations of motion take the symplectic form 
\begin{align}
\dot{\mathbf{z}} & =J^{-1}\grad H(\mathbf{z})~~,\label{def-sympl-sys}\\
J & =\begin{pmatrix}0 & I_{4\times4}\\
-I_{4\times4} & 0
\end{pmatrix}~~.
\end{align}
As a nice consequence, we can evolve the system with \textit{Gauss
Runge--Kutta schemes} which have already been shown to yield very
good results for little computational costs in previous studies, see,
e.g.,~\cite{Seyrich13,SeyrichLukes12}.

An $s$-stage Gauss Runge-Kutta scheme is a collocation method, i.e.,
an implicit Runge-Kutta scheme 
\begin{align}
\mathbf{y}_{n+1} & =\mathbf{y}_{n}+h\sum_{i=1}^{s}b_{i}f(\mathbf{Y}_{i})~~,\label{def-rk-y_n+1}\\
\mathbf{Y}_{i} & =\mathbf{y}_{n}+h\sum_{j=1}^{s}a_{ij}f(\mathbf{Y}_{j})~~,\quad i=1,...,s~~,\label{def-rk-Y_i}
\end{align}
with coefficients 
\begin{align}
a_{ij}=\int_{0}^{c_{i}}l_{j}(t)dt~~,\\
b_{j}=\int_{0}^{1}l_{i}(t)dt~~,
\end{align}
where the stages $c_{1},...,c_{s}$ are chosen as 
\begin{align}
c_{i}=\frac{1}{2}(1+\tilde{c}_{i})~~,
\end{align}
with $\tilde{c}_{i}$ being the roots of the Legendre-polynomial of
degree $s$. Here, $h$ denotes the time step size, $Y_{i}$, $i=1,...,s$,
are the so-called inner stage values and $\mathbf{y}_{n}$ denotes
the numerical approximation to the solution $\mathbf{y}$ at time
$\tau=nh$. The functions $l_{i}(t)$ are the Lagrange-polynomials
of degree $s$, 
\begin{align}
l_{i}(t)=\prod_{i\neq j}\frac{t-c_{j}}{c_{i}-c_{j}}~~.
\end{align}
Gauss Runge-Kutta methods have a convergence order $\mathcal{O}(h^{2s})$
which is the highest possible order among collocation schemes, e.g.,~\cite{hairernorsettwanner}.
Detailed information on their implementation is given in~\cite[Sec.~7]{Seyrich13},
and \cite[Chapters~VIII.5~and~VIII.6]{hairerlubichwanner}.

Very importantly, Gauss Runge--Kutta schemes almost exactly preserve
the Hamiltonian throughout the numerical evolution, cf.\cite[Fig.~10]{LSK1}.
Furthermore, it is known from numerical analysis that the solution
$\mathbf{y}_{n+1}$ coincides with the value at $t=h$ of the interpolation
polynomial $\mathbf{U}(t)$ through the points $(0,\mathbf{y}_{n})$
and $(c_{1},\mathbf{Y}_{1})...(c_{s},\mathbf{Y}_{s})$. This interpolation
polynomial can be shown to stay $\mathcal{O}(h^{s})$ close to the
exact solution of the equations of motion, see, e.g.,~ \cite[Chapter~II.1]{hairerlubichwanner}.
Therefore, we can conveniently calculate $\mathcal{O}(h^{s})$ approximations
to surface sections, such as the one presented in Fig.~\ref{Fig:KerrChaos}
above (for the procedure details see, e.g., \cite{Seyrich13}).

\section{Plunging orbits}\label{sec:PlOr}

\begin{figure}[htp]
\centerline{ \includegraphics[width=0.45\textwidth]{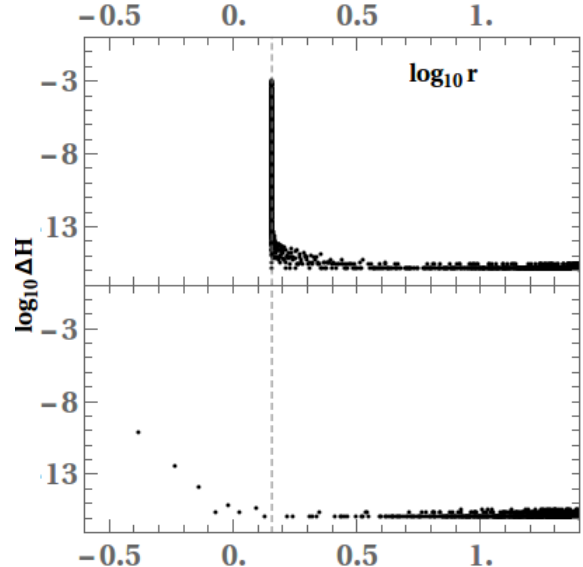}}
\caption{The relative error of the Hamiltonian $\Delta H$ as a function of the BL
radius $r$ for two plunging orbits. The top plot shows the $\Delta H$ of a plunging
orbit evolved by the Hamiltonian function discussed in section~\ref{sec:RHamBL}
(BL), while the bottom plot shows the $\Delta H$ of a plunging orbit evolved
by the Hamiltonian function discussed in section~\ref{sec:NonZamo} (KS).
In both cases we used masses $M=m=1$ and Kerr parameter $a=0.9$. For more about 
the initial conditions refer to the text. The horizon is depicted in both plots 
by a vertical dashed gray line.
}
\label{Fig:PlOr}
\end{figure}

The analytical properties of equations of motion have also an impact on the
behavior of their numerical solution. In Fig.~\ref{Fig:PlOr} we plot the
relative error of the Hamiltonian $\Delta H=|1-H(t)/H(0)|$ for plunging orbits
using the Gauss Runge-Kutta method (Appendix \ref{sec:NumMet}) with fixed time
step for equations of motion given by the Hamiltonian function discussed in 
Sec.~\ref{sec:RHamBL} (BL) (top plot) and the one discussed in section~\ref{sec:NonZamo}
(KS ) (bottom plot) as functions of the BL radius. The figure clearly shows that
the KS case covers smoothly the horizon of the Kerr black hole
(vertical dashed gray line), while the BL fails to do so by definition.

The mass parameters are $M=m=1$ and the Kerr parameter is $a=0.9$.
As initial conditions for the BL orbit we use $r=25$, $\theta=\pi/2, \phi=0$, and  
$\frac{dr}{dt}=\frac{d\phi}{dt}=\frac{d\theta}{dt}=0$. The first
two conditions for the velocities determine the initial values of $S_3$, and $P_r$.
The remaining initial conditions are $P_\phi=S_x=S_y=P_\theta=0$. The time step 
is $\Delta t_{BL}=0.1$. The initial conditions for the KS orbit are $x=25, y=z=0$,
and $P_x=P_y=P_z=0$ with spin  $S_x=S_z=0, S_z=0.9$. The time step is
$\Delta t_{KS}=0.1$.

In fact, to test how much the pole-dipole approximation of
rotating body fails near the horizon has to be investigated by other studies.
But, one can assume that not having equations of motion singular at the horizon
helps to obtain the right numerical results within the range of spins allowed
by the linear-in-spin approximation.

\end{document}